\documentclass[10pt,journal,compsoc]{IEEEtran}
\usepackage[nocompress]{cite}

\usepackage{xifthen}

\usepackage{amsmath,graphicx}
\usepackage{algorithm}
\usepackage{amsmath,amssymb,amsfonts}
\usepackage{algorithmic}
\usepackage{textcomp}
\usepackage{soul}
\usepackage{verbatim}
\def\BibTeX{{\rm B\kern-.05em{\sc i\kern-.025em b}\kern-.08em
    T\kern-.1667em\lower.7ex\hbox{E}\kern-.125emX}}
\usepackage{mathtools}
\usepackage{mathrsfs}
\usepackage{romannum}
\usepackage{graphicx}
\usepackage[flushleft]{threeparttable}

\usepackage{setspace,tabularx}
\usepackage[tight,footnotesize]{subfigure}
\usepackage{times}
\usepackage{algorithm}
\usepackage{algorithmic} 
\usepackage{amsfonts}
\usepackage{amsmath}
\usepackage{amssymb}
\usepackage{array} 
\usepackage{boxedminipage}
\usepackage{caption}
\usepackage{cite}
\usepackage{color}
\usepackage{floatflt}
\usepackage{float}
\usepackage{graphics}
\usepackage{graphicx}
\usepackage{indentfirst}
\usepackage{latexsym}
\usepackage[hidelinks]{hyperref}
\usepackage{multirow}
\usepackage{picinpar}
\usepackage{placeins}
\usepackage{psfrag}
\usepackage{subfigure}
\usepackage{theorem}
\usepackage{threeparttable}
\usepackage{wrapfig}
\usepackage{hhline}

\usepackage{todonotes}

\theoremstyle{break}

\newcommand{\ignore}[1]{ }

\newcommand{\figlbl}[1]{\label{fig.{#1}}}
\newcommand{\figref}[1]{Fig.~\ref{fig.{#1}}}

\newcommand{\seclbl}[1]{\label{sec:{#1}}}
\newcommand{\secref}[1]{Section~\ref{sec:{#1}}}
\newcommand{\eqnlbl}[1]{\label{eqn:{#1}}}
\newcommand{\eqnref}[1]{Equation~(\ref{eqn:{#1}})}

\newcommand{\beq}{\begin{equation}}
\newcommand{\eeq}{\end{equation}}

\graphicspath{{fig/}}
\begin{document}
\title{High-Speed VLSI Architectures for Modular Polynomial Multiplication via Fast Filtering and Applications to Lattice-Based Cryptography}
\author{Weihang Tan,~\IEEEmembership{Graduate Student Member,~IEEE};
        Antian Wang,~\IEEEmembership{Graduate Student Member,~IEEE};\\
        Xinmiao Zhang,~\IEEEmembership{Senior Member,~IEEE};
        Yingjie Lao,~\IEEEmembership{Senior Member,~IEEE};\\
        and~Keshab K. Parhi,~\IEEEmembership{Fellow,~IEEE}
\IEEEcompsocitemizethanks{\IEEEcompsocthanksitem 
This paper is supported in part by the Semiconductor Research Corporation under contract number 2020-HW-2998.
\IEEEcompsocthanksitem Weihang Tan and Keshab K. Parhi are with the Department of Electrical and Computer Engineering, University of Minnesota, Minneapolis, MN 55455, USA (e-mail: wtan@umn.edu; parhi@umn.edu).
\IEEEcompsocthanksitem Antian Wang and Yingjie Lao are with the Department of Electrical and Computer Engineering, Clemson University, Clemson, SC 29634, USA (e-mail: antianw@clemson.edu; ylao@clemson.edu).
\IEEEcompsocthanksitem 
 Xinmiao Zhang is  with the Department of Electrical and Computer Engineering, The Ohio State University, Columbus, OH 43210, USA (e-mail: zhang.8952@osu.edu).

}

}
\IEEEtitleabstractindextext{%
\begin{abstract}
This paper presents a low-latency hardware accelerator for modular polynomial multiplication for lattice-based post-quantum cryptography and homomorphic encryption applications. The proposed novel modular polynomial multiplier exploits the fast finite impulse response (FIR) filter architecture to reduce the computational complexity of the schoolbook modular polynomial multiplication. We also extend this structure to fast $M$-parallel architectures while achieving low-latency, high-speed, and full hardware utilization. We comprehensively evaluate the performance of the proposed architectures under various polynomial settings as well as in the Saber scheme for post-quantum cryptography as a case study. 
The experimental results show that our proposed modular polynomial multiplier reduces the computation time and area-time product, respectively, compared to the state-of-the-art designs.
\end{abstract}
}
\maketitle
\IEEEdisplaynontitleabstractindextext
\IEEEpeerreviewmaketitle
\IEEEraisesectionheading{\section{Introduction}}
\IEEEPARstart{M}{odular} polynomial multiplication is commonly used in lattice-based post-quantum cryptography (PQC) and homomorphic encryption applications. While homomorphic encryption aims at allowing computations to be directly carried out in the encrypted domain without decryption~\cite{gentry2009fully}, lattice-based cryptographic algorithms are also designed to be resistant against attacks from both traditional and quantum computers and are thus well suited for PQC. Three out of the four finalists for the NIST PQC standardization in round-3 fell into the category of lattice-based cryptography~\cite{round3}. 
In prior works, the modular polynomial multiplication for the lattice-based cryptography scheme has mostly been implemented by schoolbook polynomial multiplication~\cite{roy2020high}, number theoretic transform (NTT)~\cite{zhang2020highly} or the Karatsuba multiplication~\cite{zhang2022polynomial}. 
Different from the prior works, this paper proposes {\em novel} high-speed architectures by exploiting the fast finite impulse response (FIR) parallel filter architecture~\cite{parhi2007vlsi,parker1997low,cheng2004hardware,parker1996area,tian2021high}. The paper also proposes a novel {\em weight-stationary} systolic array for modular polynomial multiplication; these are used as building blocks for the fast parallel architecture. The proposed architecture is feed-forward and can be pipelined at arbitrary levels to achieve the desired speed. To the best of our knowledge, this is the first paper to utilize the fast parallel filter architecture to accelerate the modular polynomial multiplication for lattice-based schemes.

Exploiting the fast parallel filter approach to modular polynomial multiplication is neither straightforward nor trivial. Since the fast parallel filters contain several subfilters and merging operations, the modular operations must be incorporated at the subfilter level and merging level. No prior work has addressed these design aspects. The subfilters should correspond to single-input single-output architectures that should integrate the modular operation and should operate in real-time with no hardware under-utilization. These should also require simpler control circuits. Such designs have not been presented before.

The contributions of this paper are four-fold. First, using systolic mapping methodology~\cite{parhi2007vlsi,kung1988vlsi,jagadish1987array}, we derive a sequential weight-stationary systolic array for modular polynomial operation. This structure is {\em partly} similar to the {\em transpose-form} FIR digital filter~\cite{parhi2007vlsi} and is the main building block of the proposed architecture. The low-latency systolic array achieves full hardware utilization. Second, we propose a low-latency fast modular polynomial multiplication architecture that integrates the modular reduction at the merging level, achieves full hardware utilization and minimizes latency. Third, using {\em iterated} fast parallel filter design approach, we propose highly parallel architectures where the level of parallelism is the product of short lengths. The modular operation is also carried out at the merging step of each iteration to reduce overall latency and achieve full hardware utilization. Fourth, the advantages of the proposed architecture are demonstrated using the Saber scheme as a PQC benchmark.

The rest of the paper is organized as follows: \secref{back} reviews the mathematical background and the prior works on modular polynomial multiplication. \secref{sec_fir} and \secref{sec_fast_fir} present the details of the proposed hardware architecture, including the modular polynomial multiplier and fast $M$-parallel architecture. \secref{resu} describes the experimental results and comparisons with the state-of-the-art designs. Finally, \secref{con} concludes the paper.

\section{Background and Related Work}\seclbl{back}
In this section, we briefly review the essential notations, mathematical background, and related works. 

\subsection{Lattice-based cryptography}\seclbl{lattice}
Lattice-based cryptography relies on the NP-hard lattice problems that even quantum computers cannot solve efficiently. One example of the lattice problems is the shortest vector problem (SVP), whose security relies on the hardness of approximating SVP in the Euclidean norm~\cite{gentry2009fully}. 

There are several representative schemes for both homomorphic encryption and PQC based on the lattice-based cryptography primitive. For example, in lattice-based homomorphic encryption, BFV scheme~\cite{fan2012somewhat} and CKKS scheme~\cite{cheon2017homomorphic} support a limited number of homomorphic computations, which are also categorized as somewhat homomorphic encryption (SHE) schemes. The fully homomorphic encryption (FHE) schemes such as~\cite{gentry2009fully,gentry2013homomorphic} allow an unlimited number of homomorphic computations by using bootstrapping algorithm.

On the other hand, the NIST finalist lattice-based PQC schemes can be classified as either NTRU-based (e.g., NTRU~\cite{chen2019algorithm}), and learning with errors-based (LWE) (e.g., Crystals-Kyber~\cite{bos2018crystals} and Saber~\cite{d2018saber}) in general.

In this paper, we evaluate and compare the performance of our proposed architectures used in the Saber scheme~\cite{d2018saber} as a case study. Saber is indistinguishability under chosen-ciphertext attack secure Key Encapsulation Mechanism (KEM), which consists of three algorithms: key generation (KeyGen), encapsulation (Encaps), and decapsulation (Decaps)~\cite{d2018saber}. More specifically, the primitive of the Saber scheme is based on the hardness of the module-learning with rounding (M-LWR) problem and the use of the Fujisaki-Okamoto transform~\cite{fujisaki1999secure}. 

Among all the steps in the Saber scheme, the most widely used functions are the matrix-vector multiplication and the inner product of two vectors. For the medium-security level of Saber (post-quantum security level similar to AES-192), there are 9, 12, and 15 polynomial multiplications in the key generation, encapsulation, and decapsulation, respectively. In this paper, we only consider the medium-security level.   

Modular polynomial multiplication is a fundamental and yet the most computationally intensive operation of lattice-based cryptography. For the lattice-based PQC, modular polynomial multiplication dominates the computations across key-generation, encryption, and decryption steps in the prior works~\cite{roy2020high,mera2020compact}. Similarly, the most expensive operation for homomorphic encryption schemes is also modular polynomial multiplication. Therefore, improving the efficiency of modular polynomial multiplication is critical to the practical deployment of lattice-based PQC schemes and homomorphic encryption.

\subsection{Schoolbook modular polynomial multiplication}\seclbl{sec_schoolbook}
For the product $P(x)$ of two polynomials 
\begin{align}
    A(x) &= a[0] + a[1]x + a[2]x^2 +...+a[n-1]x^{n-1},
\\
    B(x)& = b[0] + b[1]x + b[2]x^2 +...+b[n-1]x^{n-1},
\end{align}
over $R_q$, all the coefficients of $P(x)$ need to be less than $q$ but non-negative integers, while the degree of $P(x)$ should be less than $n$, where $R_q = \mathbb{Z}_{q}/(x^n+1)$ is ring of the polynomial, and $\mathbb{Z}_{q}$ is the ring of integers modulo an integer $q$. 
The schoolbook modular polynomial multiplication between $A(x)$ and $B(x)$ modulo $(x^n+1,q)$ can be described as
\begin{align}
        &A(x) \cdot B(x) \nonumber\\
        &= \sum^{n-1}_{i = 0} \sum^{n-1}_{j = 0} a[i]b[j]x^{i+j} \text{  mod }(x^n+1,q)  \eqnlbl{eqschoolbook} \\
    &= \sum^{n-1}_{i = 0} \Big(\sum^{n-1}_{j = 0} (-1)^{\lfloor(i+j)/n\rfloor}a[i]b[j] \text{  mod  }q \Big) \cdot x^{(i+j) \text{  mod  }n }.\nonumber
\end{align}

For the schoolbook modular polynomial multiplication, the moduli are not required to be prime, which is different from the NTT-based polynomial multiplication. Consequently, the polynomial multiplication used in the M-LWR problem~\cite{alwen2013learning} and ring-learning with errors (R-LWE) problem~\cite{lyubashevsky2010ideal} can benefit from these moduli. In these cases, since all the moduli can be selected as power-of-two integers, the modular reduction for the coefficients on the schoolbook polynomial multiplication can be simply performed by keeping the least significant $\epsilon$ bits ($\epsilon$ is the bit-length of the modulus $q$, i.e., $\epsilon=\lceil\log_2(q)\rceil$) instead of using the expensive Barrett reduction~\cite{barrett1986implementing} or Montgomery modular multiplication~\cite{montgomery1985modular}. Meanwhile, schemes based on the M-LWR problem (such as the Saber scheme) obtain the error term by rounding, while naturally aligning with the power-of-two modulus. Besides, recent work shows that a power-of-two modulus can simplify and improve the polynomial multiplication for the R-LWE based homomorphic encryption schemes without affecting the computational hardness~\cite{langlois2012hardness}. The modulus in this format has been applied in some popular schemes such as BFV scheme~\cite{fan2012somewhat}. 
Based on this advantage, shortening the word-length of the operand and eliminating the modular reduction for the coefficients can increase the resource available, which can then enable the designer to increase the level of parallelism to achieve a high-speed modular polynomial multiplier. 

It may be noted that the use of the power-of-two moduli, as needed in the Saber scheme, cannot leverage the acceleration from the NTT-based polynomial multiplication without further expensive transformation. However, NTT-based polynomial multiplication has been widely applied in many lattice-based cryptography schemes when the moduli are not power-of-two~\cite{xin2020vpqc,xing2021compact,nguyen2019high,banerjee2020sapphire,zhang2020highly,nguyen2020high,roy2017hardware,tan2021pipelined}.

\subsection{Karatsuba polynomial multiplication}
To improve the efficiency and reduce the complexity of schoolbook polynomial multiplication, methods based on the divide-and-conquer strategy to increase parallelism are of great interest. {One of the examples is the Karatsuba algorithm~\cite{karatsuba1963multiplication}, which has been utilized in some prior modular polynomial multiplier designs for Saber scheme~\cite{mera2020compact,zhu2021lwrpro}.}
The 2-level Karatsuba polynomial multiplication first decomposes the input polynomials into higher-degree and lower-degree parts as $A(x) = A_0(x) + A_1(x) \cdot x^{n/2}$ and $B(x) = B_0(x) + B_1(x)\cdot x^{n/2}$ and computes 
\begin{align}
     C_0(x) &= A_0(x) \cdot B_0(x) \nonumber
     \\
     C_1(x) &= (A_0(x) + A_1(x)) \cdot (B_0(x) + B_1(x)) \nonumber\\
     C_2(x) &= A_1(x) \cdot B_1(x).
\end{align}

Then the above products are summed up, and polynomial modular reduction is carried out to derive the product $P(x)$ over the ring as
\begin{equation}
    P(x) = C_0(x) + C_3(x)\cdot x^{n/2} + C_2(x)\cdot x^n \mod (x^n+1),
\end{equation}
where 
\begin{align}
   C_3(x)= (C_1(x) - C_0(x) - C_2(x)). 
\end{align}
 Note that the degrees of $C_3(x)\cdot x^{n/2}$ and $C_2(x)\cdot x^n$ are $\frac{3}{2}n$ and $2n$, respectively. Hence polynomial subtractions are needed to perform the modular reduction by $x^n+1$. Based on this divide-and-conquer strategy of the Karatsuba algorithm, the number of coefficient multiplications is reduced from $n^2$ to $3(n/2)^2$.

\subsection{Prior hardware implementations}
Several hardware accelerators for lattice-based cryptography without using the NTT algorithm have been proposed recently~\cite{dang2019implementing,roy2020high,mera2020compact,zhu2021lwrpro,liu2019optimized,fan2018lightweight,poppelmann2014area,migliore2016hardware}. As expected, optimizing the polynomial multiplier is the main focus of these works, since it is the bottleneck. The hardware/software co-design for the modular polynomial multiplication accelerator in~\cite{dang2019implementing} shows a significant acceleration compared with the software implementation. Subsequently, the work in~\cite{mera2020compact} introduced the compact hardware/software interfacing design, which applies a hybrid method of Toom-Cook multiplication~\cite{toom1963complexity} (a generalized form of Karatsuba algorithm) and a length-64 schoolbook polynomial multiplier to optimize the modular polynomial multiplication.  A full hardware implementation is proposed in~\cite{roy2020high}, which utilizes a memory-based schoolbook polynomial multiplier. This design achieves a higher speed where each length-256 polynomial multiplication only consumes 256 clock cycles. {Later, an extended work of \cite{roy2020high} is presented in~\cite{basso2021optimized}, which is based on a design called centralized multiplier architecture. This optimized design retains the same timing performance but requires fewer hardware resources since each multiply-and-accumulate (MAC) is replaced by one multiplexer (MUX) and one adder.}
Furthermore, an 8-level recursive split hierarchical Karatsuba algorithm-based implementation is introduced in~\cite{zhu2021lwrpro}, which reduces a length-256 polynomial multiplication to only 81 clock cycles without considering the pipelining startup time. 

Besides, several architectures of modular polynomial multipliers for the R-LWE schemes are introduced in~\cite{zhang2020efficient,zhang2022polynomial,poppelmann2014area}. The works in~\cite{zhang2020efficient,poppelmann2014area} investigate the low-area design for the schoolbook modular polynomial multiplication, which only consumes fewer LUTs and DSPs. Meanwhile, the design in~\cite{zhang2022polynomial} proposes a modular polynomial multiplier using the Karatsuba algorithm and reduces the complexity by merging the polynomial modular reduction on the post-processing stage of the Karatsuba algorithm. 

However, these designs cannot consider an architecture using the fast filtering technique to reduce the latency.
Also, the architectures based on the Karatsuba algorithm generally consider the  polynomial modular reduction after the multiplication. These designs do not reduce the number of additions. Therefore, it is possible to further reduce the number of additions/subtractions at the post-processing stage, thereby reducing the total number of addition/subtraction operations. 
Since our objective is to improve the speed under a given hardware budget, we define the following two metrics in evaluating the performance from the speed perspective: 
\begin{itemize}
    \item Response time: clock cycles between the first input and the first output sample.
    \item Total latency: clock cycles between the first input and the last output sample.
\end{itemize}

\section{Modular Polynomial Multiplier based on Weight-Stationary Systolic Array}\seclbl{sec_fir}

Consider the design of a length-$n$ modular polynomial multiplier described by~\eqnref{eqschoolbook}. In this section, we use $n=4$ as an example to illustrate our proposed novel modular polynomial multiplier. 
The modular polynomial multiplication is described by:
\begin{align}
    P(x) &= A(x) \cdot B(x) \mod (x^4+1,q)\\
    &= p[0] + p[1]x+p[2]x^2 + p[3]x^3, \nonumber
    \end{align}
where 
\begin{align*}
    A(x) &= a[0] + a[1]x+a[2]x^2 + a[3]x^3,\\
    B(x) &= b[0] + b[1]x+b[2]x^2+ b[3]x^3.
\end{align*}

The polynomial multiplication of $A(x)$ and $B(x)$ leads to
\begin{align}
    P'(x) &= p'[0] + p'[1]x+p'[2]x^2 + p'[3]x^3 \nonumber \\
    &\quad+ p'[4]x^4 + p'[5]x^5 + p'[6]x^6.   \eqnlbl{p_ori}
\end{align}

Since the polynomial multiplication has a degree higher than three, the terms $x^4$, $x^5$, and $x^6$ are replaced by $-1$, $-x$, and $-x^2$, respectively, to perform the modular reduction. Thus, the coefficients of the modular polynomial multiplication are:
\begin{align}
    p[3] &= a[3] b[0] + a[2] b[1] + a[1] b[2] + a[0] b[3], \nonumber\\
    p[2] &= a[2] b[0] + a[1] b[1] + a[0] b[2] - a[3] b[3], \nonumber\\
    p[1] &= a[1] b[0] + a[0] b[1] - a[3] b[2] - a[2] b[3], \nonumber\\
    p[0] &= a[0] b[0] - a[3] b[1] - a[2] b[2] - a[1] b[3]. 
    \eqnlbl{poly_p}
\end{align}

A dependence graph (DG) of the modular polynomial multiplication for the $n=4$ example is shown in~\figref{dc_poly}.

\begin{figure}[htbp]
\centering
\resizebox{0.5\textwidth}{!}{%
\includegraphics[]{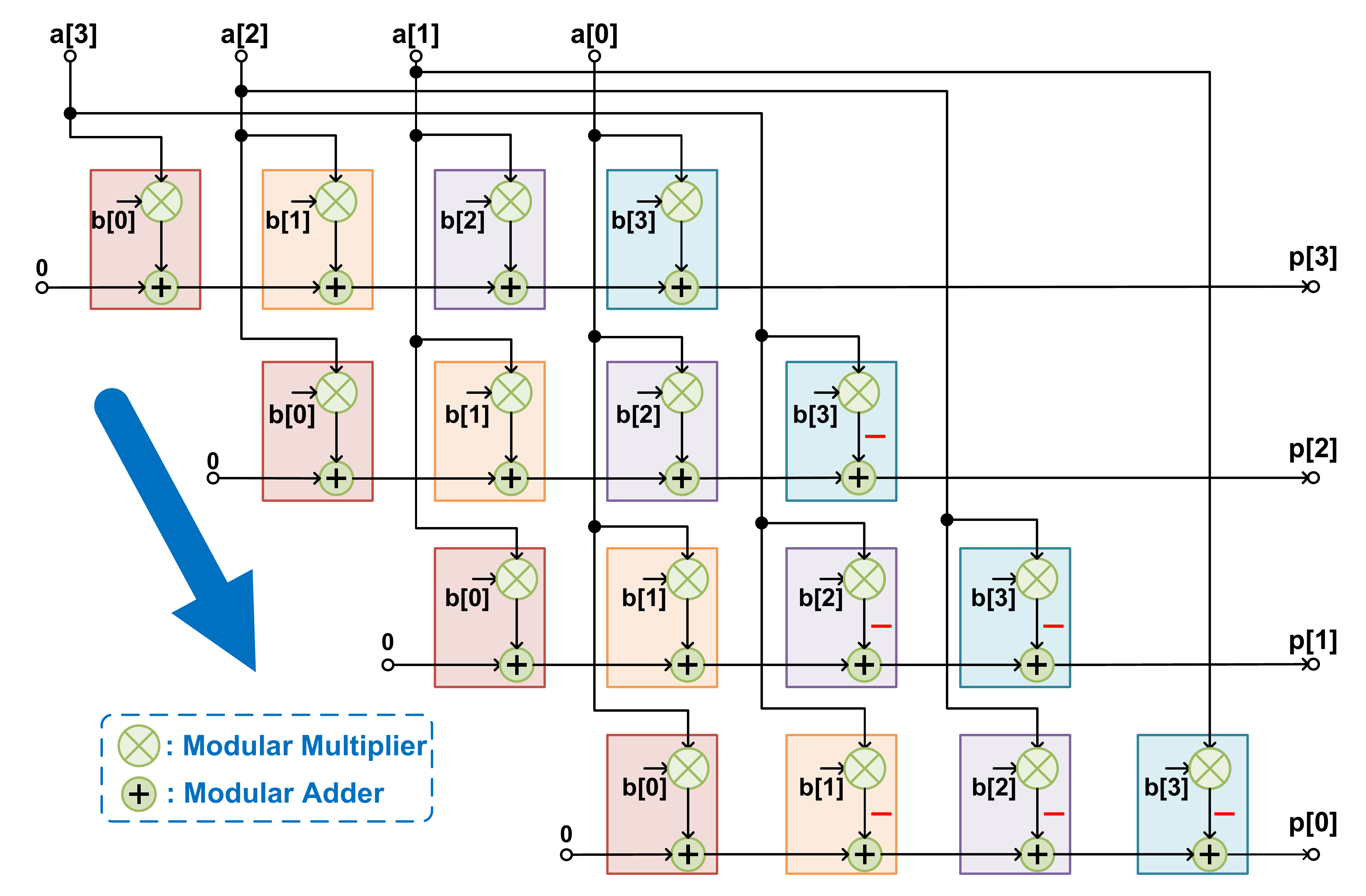}}
\caption{DG of the modular polynomial multiplication when $n=4$. The DG is mapped to a systolic array using the projection vector shown in blue.} 
\figlbl{dc_poly}
\end{figure}

 \begin{figure*}[h]
\centering
\subfigure[Direct-form]{
\resizebox{0.3\textwidth}{!}{%
\includegraphics[]{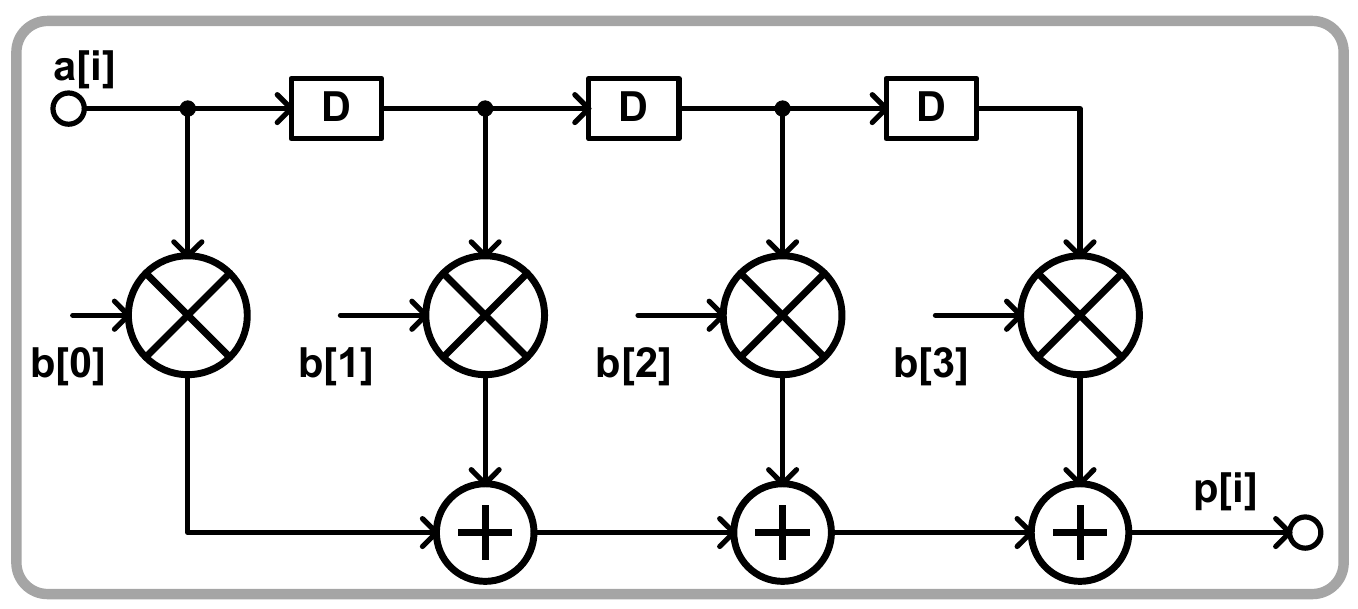}}
}
\subfigure[Transpose-form]{
\resizebox{0.3\textwidth}{!}{%
\includegraphics[]{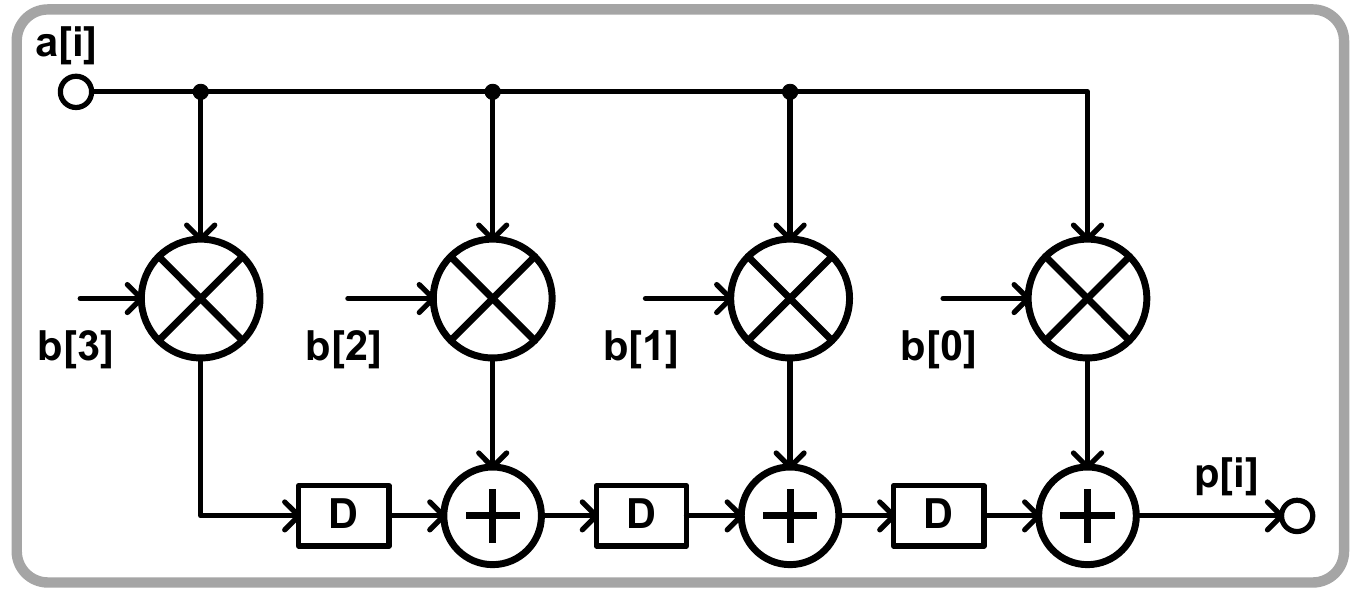}}
}
\subfigure[Hybrid-form]{
\resizebox{0.27\textwidth}{!}{%
\includegraphics[]{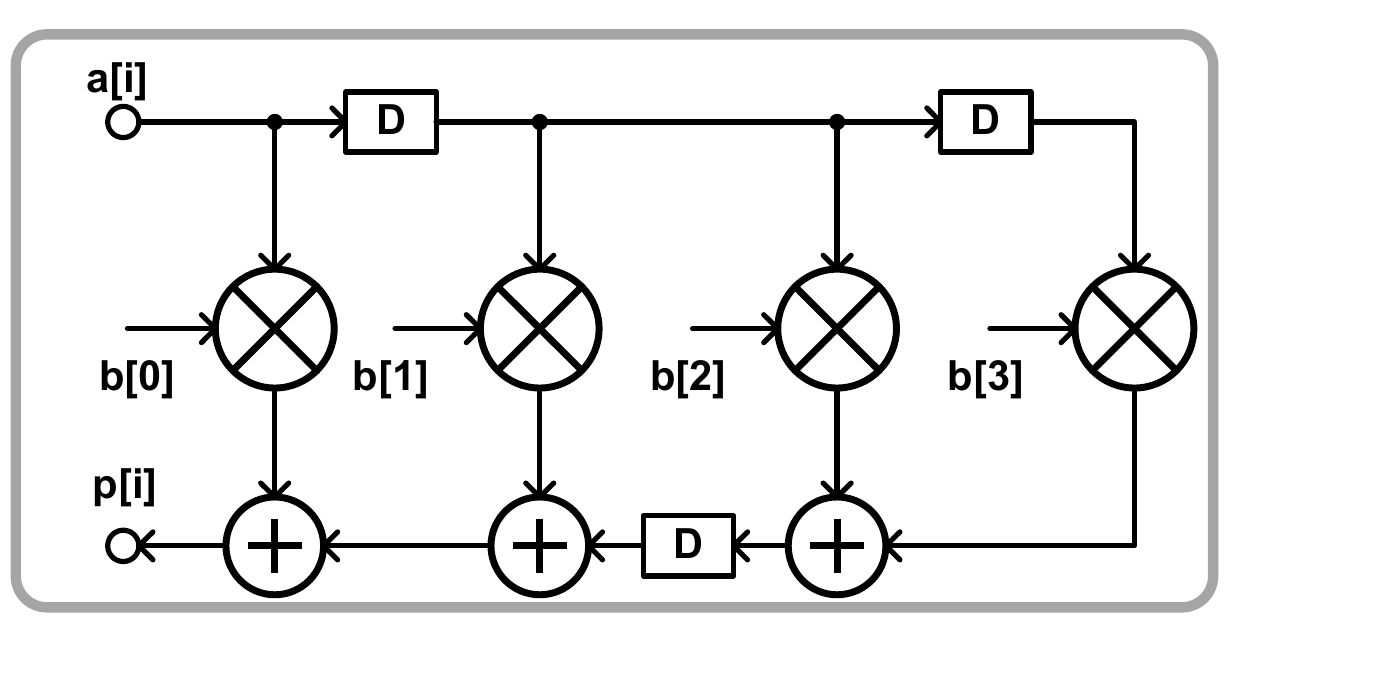}}
}
\caption{Three different forms of FIR filter architecture when $n=4$.} 
\figlbl{three_fir}
\end{figure*}

\subsection{Architecture of modular polynomial multiplier using transpose-form FIR filter}\seclbl{sec_fir_archi}
Given the similarity between modular polynomial multiplication and FIR filter, it is useful to consider three common types of FIR filter structures~\cite{parhi2007vlsi}, i.e., direct-form, transpose-form, and hybrid-form, respectively, as shown in~\figref{three_fir}. 

FIR filter is one of the  digital filters that is used to modify the frequency properties of the input signal to achieve specific design requirements~\cite{parhi2007vlsi}. It can also be mathematically expressed as a discrete convolution of two signals, which can be defined as 
\begin{equation}
    p[n] = \sum^{n-1}_{j = 0} b[j] a[n-j],
    \eqnlbl{fir}
\end{equation}
where $n$ is the number of taps, $a[n]$ is the input signal, $b[j]$ is value of the impulse response at the $j$-th instant ($j\in [0,n-1]$), and $p[n]$ is the output signal. Though using any of the FIR filter structures in~\figref{three_fir} can sufficiently instantiate \eqnref{fir} and show a negligible difference in the overall performance in most of the digital signal processing applications, the FIR structure for modular polynomial multiplication needs to be carefully selected. 

The direct-form FIR filter shown in~\figref{three_fir}(a) leads to a long critical path, which consists of one multiplier and $n$ adders. It can also be observed from \figref{three_fir}(c) that the hybrid-form architecture generates its first output immediately after loading the first input, and requires additional registers to store the intermediate results; however, the architecture is not feed-forward and has slightly longer critical path than the transpose-form. Thus, the best choice for implementing polynomial multiplication in lattice-based schemes is the transpose-form as shown in~\figref{three_fir}(b) as it has the least critical path and a feed-forward datapath. Fortunately, the DG in~\figref{dc_poly} can be mapped to a weight-stationary systolic array using the projection vector shown in blue in the figure. Alternatively, the systolic array can also be derived using the folding algorithm~\cite{parhi1992synthesis}. Note that all multiplications with coefficient $b[j]$ are mapped to the same hardware multiplier.

\begin{figure}[htbp]
\centering
\subfigure[Top-level architecture with transpose-form like structure.]{
\resizebox{0.49\textwidth}{!}{%
\includegraphics[]{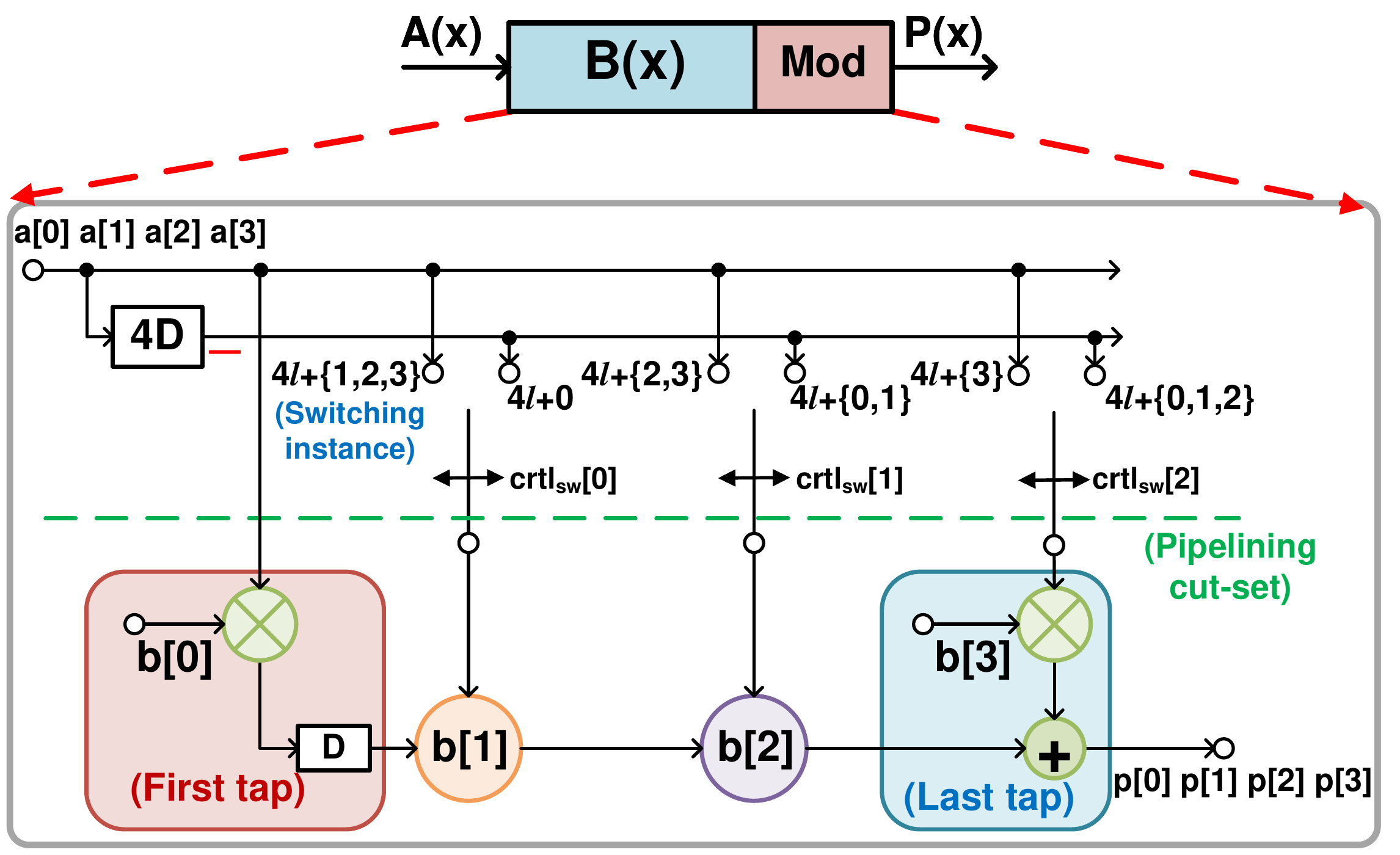}}
}
\subfigure[Details of each tap.]{
\resizebox{0.35\textwidth}{!}{%
\includegraphics[]{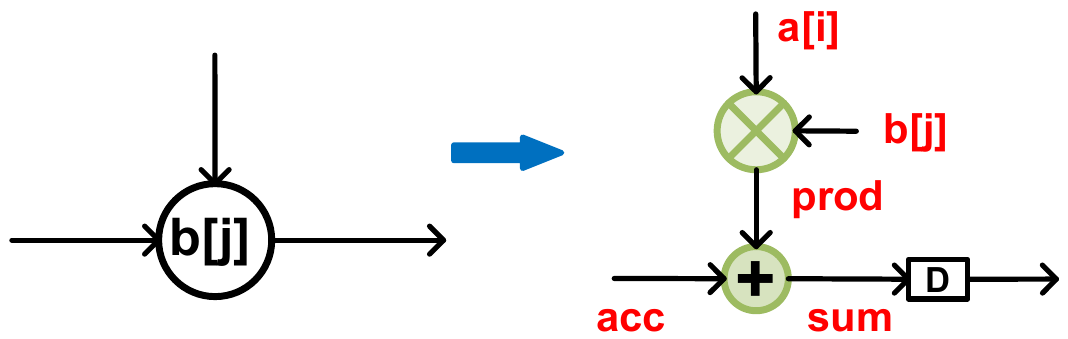}}
}
\vspace{-1em}
\caption{A degree-$4$ weight-stationary systolic modular polynomial multiplier.} 
\figlbl{sequence_archi}
\end{figure}

\figref{sequence_archi}(a) shows an example systolic architecture for modular polynomial multiplication for length-$4$ where the components in each tap (node) are illustrated in \figref{sequence_archi}(b). The systolic array contains additional switches and a shift register of size-$n$ (see the top of \figref{sequence_archi}(a)) for continuous processing of input polynomials and polynomial modular reduction. Note that using a conventional transpose-form like structure to perform the polynomial multiplication would require padding zeros until the entire operation finishes; otherwise, it will lead to conflicts and produce wrong results. Furthermore, to perform polynomial modular reduction, the shift register and switches can control the signals (coefficients of polynomial $A(x)$) properly based on the expression in \eqnref{poly_p}. 
Specifically, the coefficients of polynomial $A(x)$ with the negative signs are extracted from the shift register in its negative form. Then, the switches select either the negative form or the original form coefficients from polynomial $A(x)$ in different clock cycles. As shown in \figref{sequence_archi}, the proposed length-$4$ modular polynomial multiplier consists of four modular multipliers, three modular adders, three delay elements, three switches, and one shift register. Specifically, the shift register consists of four delay elements, and the switches are constructed using MUXs. The design in \figref{sequence_archi} can be easily extended to length-$n$. A length-$n$ modular polynomial multiplier requires $n$ modular multipliers, $(n-1)$ modular adders, $(n-1)$ delay elements, $(n-1)$ switches at the lower data paths and one shift register (consisting of $n$ delay elements). For one modular polynomial multiplication, the response time is $n$ clock cycles, while the total latency is $(2n-1)$ clock cycles. For $L$ polynomial multiplications, the response time remains the same, while the total latency in clock cycles is given by: 
\begin{equation}
    T_{lat} = n \cdot (L+1) -1.
    \eqnlbl{lat_mult}
\end{equation}
This architecture also has a full hardware utilization after the first output is computed. Hardware utilization is the percentage of the components inside this circuit that are performing useful operations, and full hardware utilization means no component is performing null operations.

The modular reduction can be performed by simply keeping the least $\epsilon$ bits for a $2^\epsilon$ modulus. For the lattice-based cryptography schemes, the degrees of the polynomial are relatively large, i.e., $n$ can be up to hundreds or thousands, which could cause a high fan-out issue on the output of the shift register and the input node. To overcome this, buffers (registers) are inserted after the switches, as shown as the green dashed line in \figref{sequence_archi}(a). As a result, the critical path is one modular multiplier and one modular adder. 

\subsection{Scheduling for the modular polynomial multiplier}\seclbl{sec_fir_ctrl}
The scheduling and control logic in the proposed architecture are very simple and efficient. The coefficients of polynomial $A(x)$ are loaded sequentially from the most significant (highest degree) coefficient to the least significant (lowest degree) coefficient while the coefficients of polynomial $B(x)$ are stored starting with the least significant coefficient to the most significant coefficient from left to right. Finally, the result coefficients are output in the same order as $A(x)$ (i.e., from the most significant coefficient to the least significant coefficient).

 The notation {$4l+\{0,1,2,3\}$ represents the switch instances with a switch period of 4 clock cycles. Hence, $l$ can be interpreted as the $l$-th period (iteration). For example, the left node will be connected when the switch instances are $4l+ \{1, 2,3\}$, while the right node will be connected at the switch instance of $4l+0$. }
Each switch is controlled by a one-bit signal from the $(n-1)$-bit controller $ctrl_{sw}$: if this bit is equal to $1$, the operand from polynomial $A(x)$ of the modular multiplier is loaded from the input node; otherwise, it is loaded from the shift register. These control signals $ctrl_{sw}$ can be simply generated by a counter (ranging from $0$ to $(n-2)$), as:
\begin{equation}
   ctrl_{sw} =  \left\{\begin{array}{ll}
    \{0,...,0,0\}, & \text{ if counter} = 0, \\
    \{ctrl_{sw}[n-3:0],1\}, & \text { otherwise. }
\end{array}\right.
\end{equation}
After resetting the counter, all $(n-1)$-bit control signals $ctrl_{sw}$ are zeros. Then, in every subsequent clock cycle, $ctrl_{sw}$ shifts left by padding a ``$1$'' to the least significant bit (LSB).

\section{Fast Polynomial Multiplier Using Fast M-Parallel Filter Architecture}\seclbl{sec_fast_fir}

In this section, we derive a highly parallel hardware architecture for the polynomial multiplication based on the fast parallel filter algorithm~\cite{parhi2007vlsi,parker1997low,parker1996area,cheng2004hardware}. 
The proposed design requires less resource overhead than prior Karatsuba-based polynomial multipliers in the post-processing stage. Parallel structures for modular polynomial multiplication for small lengths are first derived. These can then be {\em iterated} to obtain architectures for larger levels of parallelism. For example, a fast $2$-parallel (i.e., $M=2$) modular polynomial multiplier can be iterated twice (or thrice) to design a $4$-parallel (or $8$-parallel) multiplier.

\subsection{Fast 2-parallel architecture}\seclbl{sec_fast_2}
The fast $2$-parallel modular polynomial multiplication, referred to as Fast.2.PolyMult, is described in Algorithm~\ref{fasttwo}, which mainly consists of three stages: pre-processing (Step 1), intermediate polynomial multiplication (Step 2), and post-processing (Steps~3~and~4).

We first decompose the polynomials $A(x)$ and $B(x)$ based on the even and odd indices, as shown in Step 1, also called polyphase decomposition~\cite{parhi2007vlsi}. We denote $y = x^2$, and the polynomial $A(x)$ is expressed as:
\begin{equation}
A(x) = A_0(x^2) + A_1(x^2)\cdot x 
    = A_0(y) + A_1(y)\cdot x, 
    \eqnlbl{a_fast_two}
\end{equation}
where the even indexed polynomial $A_0(y)$ and  the odd indexed polynomial $A_1(y)$ are expressed as:
\begin{align}
    A_0(y) &=  a[0] + a[2]y+a[4]y^2+
    \hdots\nonumber \\
    & +a[n-2]y^{n/2-1} \mod(y^{n/2}+1), \\
    A_1(y)  &=  a[1] + a[3]y+a[5]y^2+ \hdots\nonumber\\
    & +a[n-1]y^{n/2-1} \mod(y^{n/2}+1).
\end{align}
A similar decomposition is applied to $B(x)$ to obtain its even indexed and odd indexed polynomials $B_0(y)$ and $B_1(y)$. The product $P(x)$ can be computed as:
\begin{align}
&P(x)= P_{0}\left(y\right)+  P_{1}\left(y\right)\cdot x \nonumber\\
=&\left(A_{0}(y) + A_{1}(y)\cdot x\right) \cdot\left(B_{0}(y) + B_{1}(y)\cdot x\right) \nonumber\\
=& A_{0}\left(y\right) B_{0}\left(y\right)+ \left[A_{0}\left(y\right) B_{1}\left(y\right)+A_{1}\left(y\right) B_{0}\left(y\right)\right]\cdot  x\nonumber\\
&+\left[A_{1}\left(y\right) B_{1}\left(y\right)\right]\cdot y
\end{align}
\begin{algorithm}[htbp]
\caption{\textbf{Fast.2.PolyMult($A(x),B(x)$)}}

\label{fasttwo}
\hspace*{\algorithmicindent}
\textbf{Input:}  $A(x)$ and $B(x)$ $\in R_q $

\hspace*{\algorithmicindent} \textbf{Output:} $P(x) = (P_0(x^2),P_1(x^2))$

\hspace*{\algorithmicindent}
\hphantom{\textbf{Oput::}}
//$P(x)=A(x) \cdot B(x) \mod (x^n+1,q)$

\begin{algorithmic}[1]

    \STATE $A(x) = A_{0}(x^2) + A_{1}(x^2)\cdot x$
    
     //Split $A(x)$ as two parts based on odd and even indices 

     $B(x) = B_{0}(x^2) + B_{1}(x^2)\cdot x$
     
     //Split $B(x)$ as two parts based on odd and even indices 

    \STATE $U(y) = A_{0}(y)B_{0}(y) \mod (y^{n/2}+1,q)$, where $y = x^2$

     $V(y) = A_{1}(y)B_{1}(y) \mod (y^{n/2}+1,q)$ 

     $W(y) = (A_{0}(y)+A_{1}(y))(B_{0}(y)+B_{1}(y))$ \\
     $\quad \quad \quad\mod (y^{n/2}+1,q)$ 
     
// Intermediate modular polynomial multiplication 
    \STATE $P_0(y) = U(y)+V(y)\cdot y \mod (y^{n/2}+1,q)$  $\quad$ \\
    $P_1(y) = W(y)-(U(y)+V(y))  \mod (y^{n/2}+1,q)$ 
    \STATE $P(x) = P_0(x^2)+P_1(x^2)\cdot x$, where $y = x^2$ 
       \RETURN {$P(x)$}

\end{algorithmic}
\end{algorithm}
The polyphase decomposition describes one polynomial multiplication of length-$n$ in terms of four polynomial multiplications of length-$n/2$. While this step in itself does not reduce the computation complexity, it is an essential first step. In Step 2, the fast filter algorithm describes the modular polynomial multiplication in terms of three polynomial multiplications of half length; this reduces the complexity by 25\%. Denote the three intermediate modular polynomial multiplication outputs as $U(y)$, $V(y)$, and $W(y)$. In the fast algorithm, $P_1(y)$ is computed as:
\begin{align}
    P_1\left(y\right)&=A_0\left(y\right)B_1\left(y\right)+A_1\left(y\right)B_0\left(y\right) \nonumber\\
    &= \left(A_0\left(y\right)+A_1\left(y\right)\right)\left(B_0\left(y\right)+B_1\left(y\right)\right) \nonumber\\
    &\quad-A_0\left(y\right)B_0\left(y\right)-A_1\left(y\right)B_1\left(y\right)\\
    & = W(y)-(U(y)+V(y)),
\end{align}
where 
\vspace{-1em}
\begin{align}
    U(y) &= A_0(y)B_0(y),\\
    V(y) &= A_1(y)B_1(y),\\
    W(y) &= (A_0(y)+A_1(y))(B_0(y)+B_1(y)).
\end{align}

Note that unlike $P_1(y)$, $P_0(y) = U(y)+V(y)\cdot y$ mod $(y^{n/2}~+~1)$ requires further modular polynomial reduction, which is achieved in the post-processing step.
Since $V(y)$ needs to be multiplied by $y$ before adding the coefficients of $U(y)$, the highest degree of coefficient exceeds the range of the ring ($y^{n/2}+1$), (i.e., $U(y)+V(y)\cdot y= u[0] + p_0[1]y+p_0[2]y^2+...+v[n/2-1]y^{n/2}$). 
As a result, the even polynomial $P_0(y)$ requires an additional subtraction and is computed as: 
\begin{align}
   P_0(y) =&(u[0]-v[n/2-1])+ p_0[1]y+p_0[2]y^2\nonumber \\
   &+\hdots+ p_0[n/2-1]y^{n/2-1}.
   \eqnlbl{p0}
\end{align}

The data-flow of the proposed fast parallel architecture is shown in \figref{fast_two_flow}.
\begin{figure}[htbp]
\centering
\resizebox{0.4\textwidth}{!}{%
\includegraphics[]{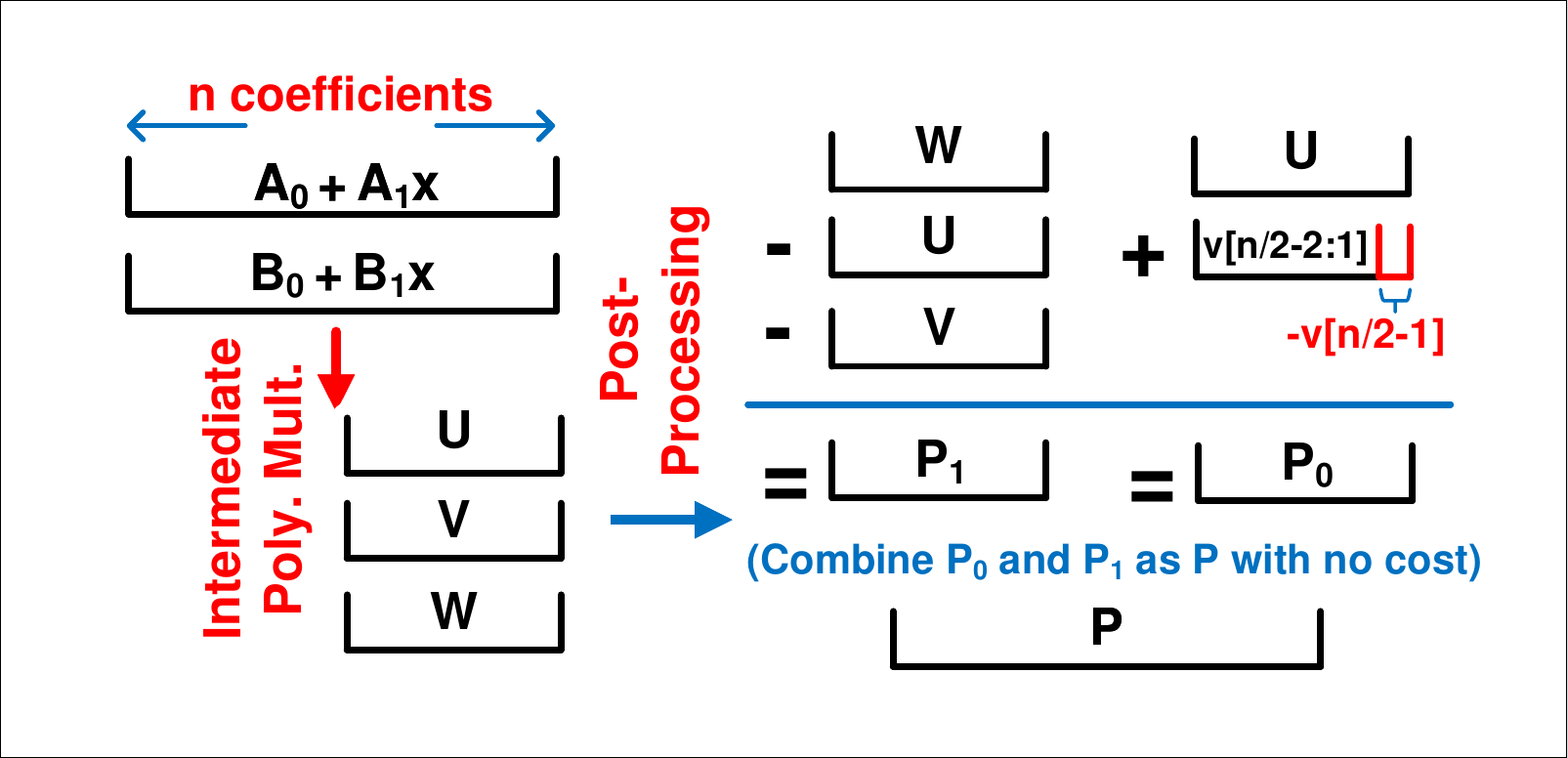}}
\caption{Data-flow of the Fast.2.PolyMult algorithm.}
\figlbl{fast_two_flow}
\end{figure}

Different from the traditional methods that execute the polynomial modular reduction during or after post-processing (i.e., combining the intermediate polynomials back to a single polynomial)~\cite{mera2020compact,zhang2022polynomial}, we integrate polynomial modular reduction into the three intermediate polynomial multiplications. This is achieved by using the sequential systolic modular polynomial multiplication described in the previous section. A 2-level Karatsuba polynomial multiplication requires at least $(n-1)$ clock cycles to output $n$ coefficients sequentially for the three intermediate polynomials and $(\frac{7}{2}n-4)$ or $(3n-3)$ modular additions/subtractions for post-processing~\cite{zhang2022polynomial}. 
In contrast, by employing the sequential weight-stationary systolic polynomial modular multiplier as shown in \figref{sequence_archi}, $\frac{n}{2}$ coefficients of $U(y)$, $V(y)$, and $W(y)$ are output in the same $(n-1)$ clock cycles without requiring additional elements. As these three intermediate polynomials are already in the ring $R_q$, the post-processing stage has a lower cost, which only needs $\frac{3}{2}n$ modular additions/subtractions.

\figref{fast_two} depicts the proposed hardware architecture for Algorithm~\ref{fasttwo} for a length-$n$ modular polynomial multiplication. It mainly consists of four adders/subtractors, three registers, and three length-$\frac{n}{2}$ modular polynomial multipliers that also include three shift registers of size-$\frac{n}{2}$ (as described in \figref{sequence_archi}). 
Besides, the bottom path can store $v[n/2-1]$ (coefficient from $V(y)$) for $\frac{n}{2}$ clock cycles and feed its negative form ($-v[n/2-1]$) to the adder at the upper path in each iteration $l$. This operation is controlled by two switches. When the left switch's instance is at $(n/2)l+0$, the output coefficient of $V(y)$ is loaded into a register, while the right switch will release the stored data to the next operation at $(n/2)l+(n/2-1)$. 
\begin{figure}[htbp]
\centering
\resizebox{0.48\textwidth}{!}{%
\includegraphics[]{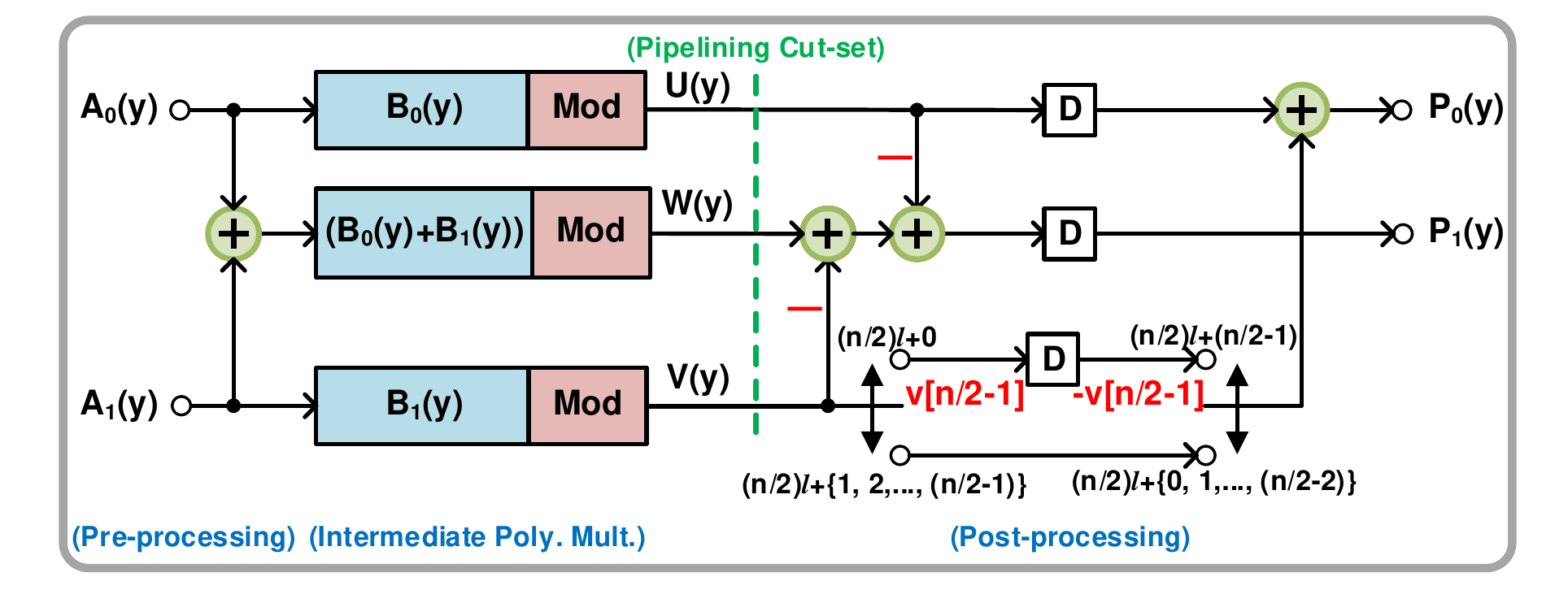}}
\caption{Fast $2$-parallel modular polynomial multiplier.}
\figlbl{fast_two}
\end{figure}

The coefficients of $P_1(y)$ can be simply obtained by using two subtractors, while the coefficients for $P_0(y)$ are more complicated to generate. The addition between $U(y)$ and $V(y) \cdot y$ is explained using the timing diagrams for $n=8$ shown in \figref{fast_two_illu}. As the coefficients of $U(y)$ and $V(y)$ are generated in the same pattern as shown in \figref{fast_two_illu}(a), directly calculating $P_0(y)$ is infeasible without multiplying $y$ for $V(y)$. However, delaying $U(y)$ by one cycle can enable the addition operation as shown in \figref{fast_two_illu}(b). Furthermore, to perform the polynomial modular reduction as in \eqnref{p0}, as described in \figref{fast_two_illu}(c), two switches and two delay elements are required. For the subtraction of $v[3]$ from $u[0]$, the first switch passes $v[3]$ to the delay element and the second switch releases its negative after four clock cycles ($\frac{n}{2}$ clock cycles for general case), as $u[0]$ is output four clock cycles ($\frac{n}{2}$ clock cycles for general case) after $v[3]$. Note that no additional adder/subtractor is needed and full hardware utilization is retained for all the components in the circuit. Moreover, this optimization technique still allows continuous processing of modular polynomial multiplications without requiring any null operations. To align the coefficients of $P_1(y)$ with $P_0(y)$, one delay element is placed at the end of $P_1(y)$'s output. 

\begin{figure}[htbp]
\centering
\resizebox{0.48\textwidth}{!}{%
\includegraphics[]{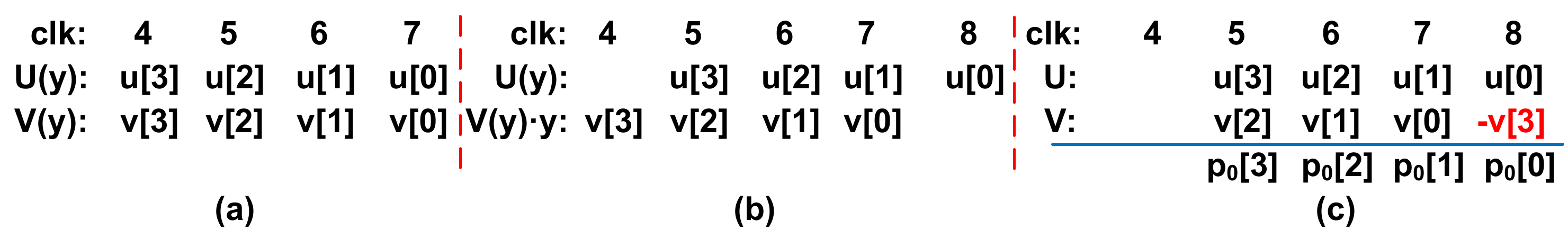}}
\caption{Timing diagram for $P_0[y]$ at post-processing stage when $n=8$. }
\figlbl{fast_two_illu}
\end{figure}

While the fast modular polynomial multiplier structure is similar to the fast parallel filter, there is one fundamental difference. Unlike the fast parallel filter where all computations are causal, the computation $ V(y)\cdot y$ is inherently a {\em non-causal} operation. This is transformed into a causal operation by introducing a latency of one clock cycle; this can be achieved by placing delays at one feed-forward cut-set in the post-processing step. The proposed {\em novel} approach of computing $ V(y)\cdot y$ does not increase the latency beyond one clock cycle and preserves the feed-forward property of the architecture and continuous data-flow property.

\subsection{Fast 4-parallel architecture}
A fast $4$-parallel architecture can be derived by iterating the fast $2$-parallel architecture twice~\cite{parhi2007vlsi,parker1997low,parker1996area,cheng2004hardware}. The fast $4$-parallel schoolbook modular polynomial multiplication algorithm (also denoted as Fast.4.PolyMult) is presented in Algorithm \ref{fastfour}, while \figref{fast_four} shows the corresponding architecture. 

\begin{algorithm}[htbp]
\caption{\textbf{Fast.4.PolyMult($A(x),B(x)$)}}
\label{fastfour}
\hspace*{\algorithmicindent}
\textbf{Input:}  $A(x)$ and $B(x)$ $\in R_q$

\hspace*{\algorithmicindent}
\textbf{Output:} $P(x) = (P_0(x^4),P_1(x^4),P_2(x^4),P_3(x^4))$,

\hspace*{\algorithmicindent}\hphantom{\textbf{Output:}}//$P(x) = A(x) \cdot B(x) \mod (x^n+1,q)$

\begin{algorithmic}[1]

    \STATE $A(x) = A_{0}(x^2) + A_{1}(x^2)\cdot x^2$
    
    //Split $A(x)$ as two parts based on odd and even indices 

     $B(x) = B_{0}(x^2) + B_{1}(x^2)\cdot x^2$
     
     //Split $B(x)$ as two parts based on odd and even indices  
     
    \STATE 
    ($C_0(y),C_1(y)$) = Fast.2.PolyMult($A_0(x^2), B_0(x^2)$), 
    
    where $y = x^4$
    
    ($C_2(y),C_3(y)$) = Fast.2.PolyMult$\Big((A_0(x^2)+A_1(x^2))$,
    
    $(B_0(x^2)+B_1(x^2))\Big)$\\
    ($C_4(y),C_5(y)$) = Fast.2.PolyMult($A_1(x^2), B_1(x^2)$)
    
    \STATE $P_0(y) = C_0(y)+C_5(y)\cdot y\mod (y^{n/4}+1,q)$ \\
           $P_1(y) = C_2(y)-C_1(y)-C_4(y)  \mod (y^{n/4}+1,q) $\\
           $P_2(y) = C_1(y)+C_4(y)  \mod (y^{n/4}+1,q) $\\
           $P_3(y) = C_3(y)-C_0(y)-C_5(y)  \mod (y^{n/4}+1,q) $
    \STATE $P(x) = P_0(x^4)+P_1(x^4)\cdot x + P_2(x^4)\cdot x^2 +  P_3(x^4)\cdot x^3 $, where $y = x^4$
       \RETURN {$P(x)$}

\end{algorithmic}
\end{algorithm}
\begin{figure}[htbp]
\vspace{-1em}
\centering
\resizebox{0.48\textwidth}{!}{%
\includegraphics[]{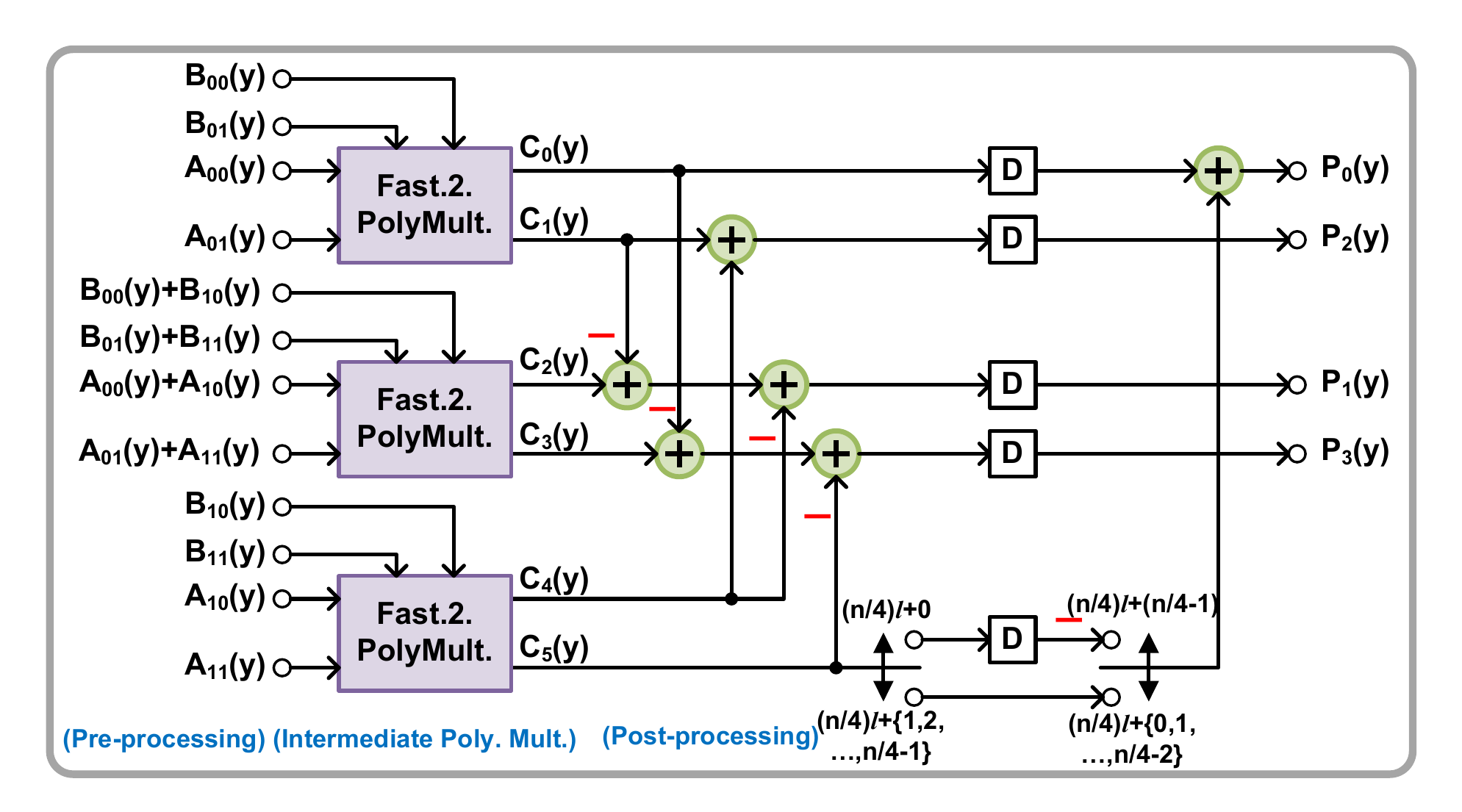}}
\caption{Fast $4$-parallel modular polynomial multiplier.}
\figlbl{fast_four}
\end{figure}

The Fast.4.PolyMult algorithm has four steps. In Step 1 of Algorithm \ref{fastfour}, $A(x)$ is split into two parts based on the odd and even indices. Then, $A_0(x^2)$, $A_1(x^2)$, and their sum $(A_0(x^2) + A_1(x^2))$ are further split based on Step 1 in Fast.2.PolyMult (Algorithm \ref{fasttwo}). $A_0(x^2)$ and $A_1(x^2)$ are decomposed as four polynomials $(A_{00}(x^4), A_{01}(x^4), A_{10}(x^4), A_{11}(x^4))$ which are fed to upper and lower fast $2$-parallel modular polynomial multipliers (denoted Fast.2.PolyMult. in \figref{fast_four}), respectively. Meanwhile, as the fast $2$-parallel modular polynomial multiplier has two inputs in parallel, $(A_0(x^2)+A_1(x^2))$ in Step 2 is simply implemented as two adders in the middle fast $2$-parallel modular polynomial multiplier, i.e., $(A_{00}(x^4) + A_{10}(x^4))$ and $(A_{01}(x^4)+A_{11}(x^4))$. 
Let $y$ represent $x^4$; Hence the four polynomials decomposed from $A_0(x^2)$ and $A_1(x^2)$ can be expressed as 
\begin{align}
    A_{00}(y) &= a[0] + a[4]y + a[8]y^2 + ...\nonumber \\
    & \quad+ a[n-4]y^{n/4-1} \mod(y^{n/4}+1),\\
    A_{10}(y) &= a[1] + a[5]y + a[9]y^2 + ... \nonumber\\
    & \quad+ a[n-3]y^{n/4-1} \mod(y^{n/4}+1),
\\
    A_{01}(y) &= a[2] + a[6]y + a[10]y^2 + ... \nonumber\\
    & \quad+ a[n-2]y^{n/4-1} \mod(y^{n/4}+1),\\
    A_{11}(y) &= a[3] + a[7]y + a[11]y^2 + ... \nonumber\\
    & \quad+ a[n-1]y^{n/4-1} \mod(y^{n/4}+1),
\end{align}
where 
\vspace{-0.5em}
\begin{equation}
    A(x) = A_{00}(x^4) + A_{10}(x^4) \cdot x+  A_{01}(x^4) \cdot x^2 + A_{11}(x^4) \cdot x^3.
\end{equation}
$B(x)$ can be decomposed in a similar manner. 

In the intermediate polynomial multiplication stage, three length-$\frac{n}{4}$ fast $2$-parallel modular polynomial multipliers (\figref{fast_two}) generate six length-$\frac{n}{4}$ polynomials, $C_0(y), C_1(y),\hdots,C_5(y)$. As shown in \figref{fast_four}, $\frac{3}{2}n$ additions/subtractions are carried out by six adders/subtractors, where each adder/subtractor performs $\frac{n}{4}$ additions/subtractions. Finally, polynomial modular reduction for $C_5(y)$ and $C_0(y)$ are performed in a manner similar to the fast $2$-parallel architecture (\figref{fast_two}).

\subsection{Fast 3-parallel architecture}

We also present the design for a fast $3$-parallel schoolbook modular polynomial multiplication algorithm (denoted as Fast.3.PolyMult), allowing $M$ to be a multiple of 3, enabling various levels of parallelism. Fast.3.PolyMult algorithm also consists of three stages, which is illustrated in Algorithm \ref{fastthree}. During the polyphase decomposition (pre-processing stage), polynomial $A(x)$ is decomposed as 
\begin{equation}
            A(x) = A_0(x^3) + A_1(x^3)\cdot x +  A_2(x^3)\cdot x^2.
            \eqnlbl{a_three}
\end{equation}
The modular multiplication result $P(x)$ can be defined as:
\begin{equation}
    P(x) = P_0(y) + P_1(y)\cdot x + P_2(y) \cdot x^2,
\end{equation}
where $y = x^3$, and these three sub-polynomials are presented in Step 4 in Algorithm~\ref{fastthree}. The derivation of the fast $3$-parallel modular multiplier is similar to the fast parallel filter derivation; the reader is referred to \cite{parhi2007vlsi,parker1997low,parker1996area}.
\begin{algorithm}[htbp]
\caption{\textbf{Fast.3.PolyMult($A(x),B(x)$)}}
\label{fastthree}

\hspace*{\algorithmicindent}
\textbf{Input:} $A(x)$ and $B(x)$ $\in R_q$

\hspace*{\algorithmicindent}
\textbf{Output:} $P(x) = (P_0(x^3),P_1(x^3),P_2(x^3))$,

\hspace*{\algorithmicindent}\hphantom{\textbf{Output:}}//$P(x)=A(x) \cdot B(x) \mod (x^n+1,q)$

\begin{algorithmic}[1]
    \STATE $A(x) = A_{0}(x^3) + A_{1}(x^3)\cdot x + A_{2}(x^3)\cdot x^2$  
   
     $B(x) = B_{0}(x^3) + B_{1}(x^3)\cdot x +  B_{2}(x^3)\cdot x^2$

    \STATE 
    $C_0(y) = A_0(y)B_0(y) \mod (y^{n/3}+1,q)$
    
    $C_1(y) = A_1(y)B_1(y) \mod (y^{n/3}+1,q)$
    
    $C_2(y) = A_2(y)B_2(y) \mod (y^{n/3}+1,q)$
    
    $C_3(y) = \big(A_0(y)+A_1(y)\big) \big(B_0(y)+B_1(y)\big)$\\
    $\quad \quad \quad \mod (y^{n/3}+1,q)$
    
    $C_4(y) = \big(A_1(y) + A_2(y)\big)\big(B_1(y) + B_2(y)\big)$ \\$\quad \quad \quad \mod (y^{n/3}+1,q)$
    
    $C_5(y) = \big(A_0(y)+A_1(y) + A_2(y)\big) \big(B_0(y)+B_1(y) + B_2(y)\big) \mod (y^{n/3}+1,q)$, where $y = x^3$
    \STATE 
    $D_0(y) = C_3(y) - C_1(y) \mod (y^{n/3}+1,q)$
    
    $D_1(y) = C_4(y) - C_1(y) \mod (y^{n/3}+1,q)$
    
    $D_2(y) = C_0(y) - C_2(y) \cdot y \mod (y^{n/3}+1,q)$
    
    $D_3(y) = C_5(y) \mod (y^{n/3}+1,q)$
    \STATE 
     $P_0(y)  = D_2(y) + D_1(y) \cdot y\mod(y^{n/3}+1,q)$
     
     $P_1(y)  = D_0(y) - D_2(y) \mod (y^{n/3}+1,q)$
     
     $P_2(y)  = D_3(y) - D_0(y) - D_1(y) \mod (y^{n/3}+1,q)$
     
    \STATE $P(x) = P_0(x^3)+P_1(x^3)\cdot x + P_2(x^3)\cdot x^2$,
 where $y = x^3$
      \RETURN {$P(x)$}
\end{algorithmic}
\end{algorithm}
\begin{figure}[htbp]
\centering
\resizebox{0.48\textwidth}{!}{
\includegraphics[]{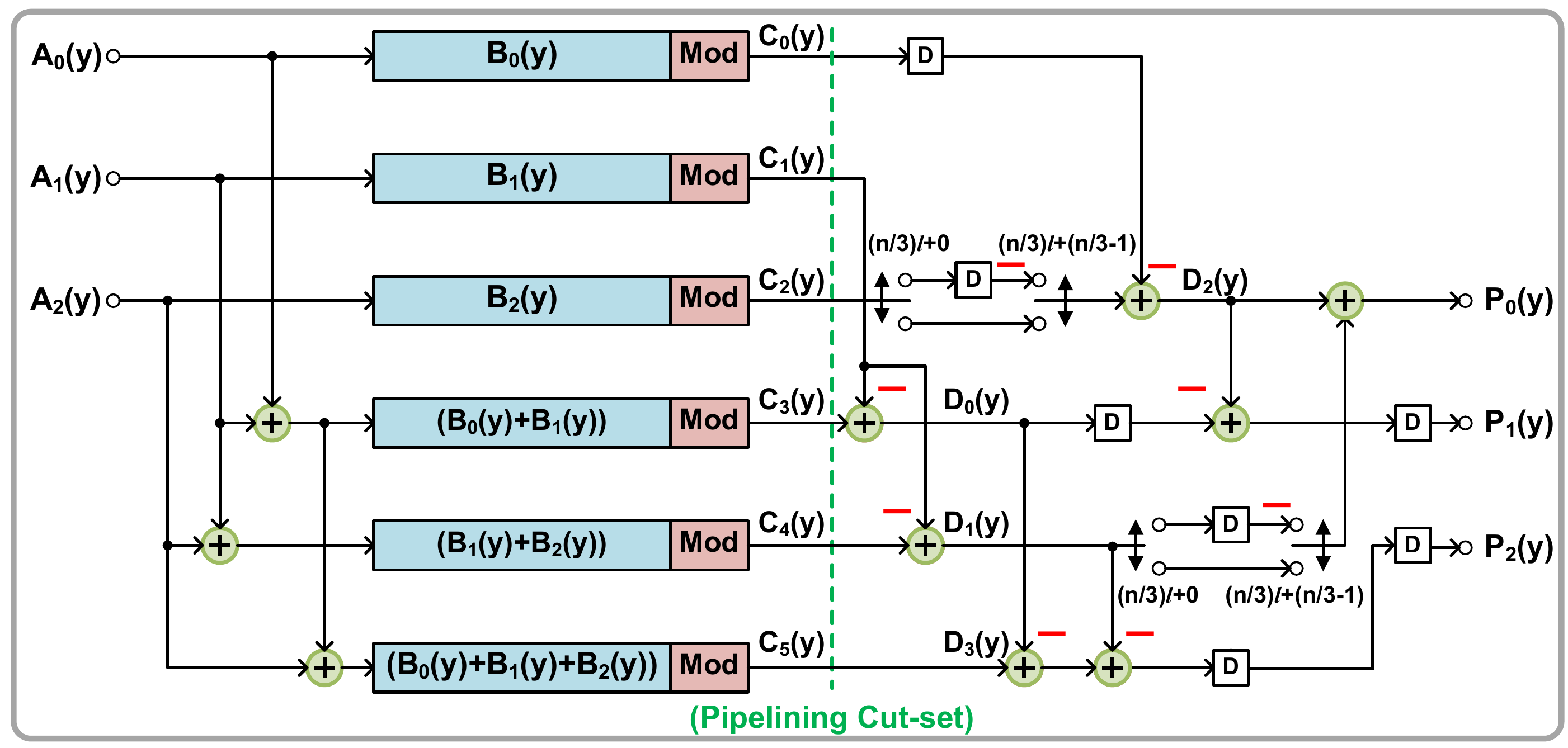}}
\vspace{0.5em}
\caption{Fast $3$-parallel modular polynomial multiplier.}
\figlbl{fast_three}
\end{figure}

The architecture for the Fast.3.PolyMult algorithm is shown in \figref{fast_three}, which consists of six length-$\frac{n}{3}$ modular polynomial multipliers, thirteen modular adders/subtractors with additional delay elements. These six length-$\frac{n}{3}$ modular multipliers compute the intermediate polynomials $C_0(y)$ to $C_5(y)$ with an additional pipelining stage at the end of the modular multipliers' output. 

In the post-processing stage, six intermediate polynomials are used to generate four new intermediate polynomials $D_0(y)$ to $D_3(y)$ before computing the outputs $P_0(y)$, $P_1(y)$, and $P_2(y)$ using fewer additions/subtractions.

\subsection{Fast M-parallel architecture}\label{sec_fast_m}
Using the {\em iterated} approach, we can use fast $2$-parallel architecture and/or fast $3$-parallel architecture to achieve higher levels of parallelism. Therefore, we can implement various fast $M$-parallel architectures, where the level of parallelism $M$ can be a power-of-two integer, power-of-three integer, or product of any power-of-two and power-of-three. 
\begin{figure}[htbp]
\centering
\resizebox{0.40\textwidth}{!}{%
\includegraphics[]{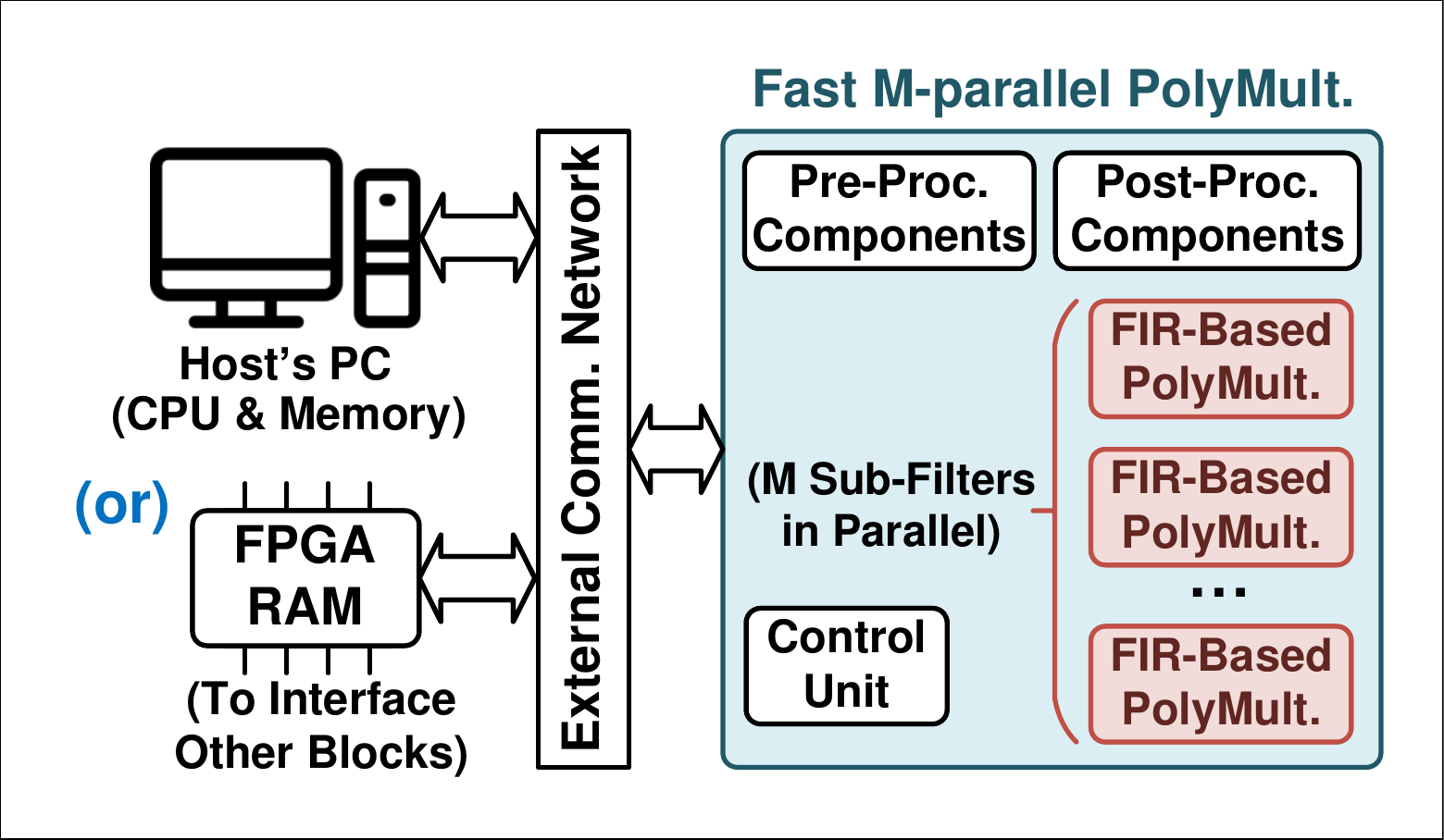}}
\caption{High-level overview of generalized fast $M$-parallel modular polynomial multiplier.}
\figlbl{fast_m}
\end{figure}

The high-level overview of the generalized fast $M$-parallel architecture is shown in \figref{fast_m}. This architecture mainly has $M$ sub-modular polynomial multipliers of length-$\frac{n}{M}$ operating in parallel to generate $M$ sub-polynomials of $P(x)$. In addition, the components for post-processing as well as the control unit, are used to align the coefficients from all the output sub-polynomials of $P(x)$. This is similar to inserting a pipelining cut-set to transform non-causal operations into causal operations, at the expense of an increase in latency by one cycle. During the computation, the data can be either accessed from the host's personal computer (PC) to get the input polynomials' coefficients directly or communicated to the FPGA's RAM.

The timing performance can be theoretically derived as follows. The fast $M$-parallel design can reduce the response time to approximately $n/M$ clock cycles.
In general, the total latency of an $M$-parallel modular polynomial multiplier for $L$ polynomial multiplications can be expressed as: 
\begin{equation}
    T_{lat} = n(1+L)/M + \lceil \log_2(M)\rceil.
\end{equation}

\section{Experimental Results}\seclbl{resu}
\begin{table*}[htbp]
  \centering
  \caption{Comparison of area consumption and frequency for modular polynomial multiplier when $n=256$}\label{tb_mul}
  {
\begin{tabular}{|c||c|c|c|c|c|c|}
\hline
Design &Device &LUTs &FFs &DSPs &BRAM &Freq. [MHz] \\  \hline 
Roy (1 Mult.)~\cite{roy2020high} &Ultrascale+  &17406 &5083 &0 &0 &250  \\  \hline 
Roy (2 Mult.)~\cite{roy2020high} &Ultrascale+  &31853 &8844 &0 &0 &250  \\  \hline 
Zhu~\cite{zhu2021lwrpro} &Ultrascale+  &13954 &3943 &85 &6 & 100 \\ \hline
Basso (HS-I 256)~\cite{basso2021optimized} &Ultrascale+  &10844 &5150 &0 &0 &250  \\  \hline 
Basso (HS-I 512)~\cite{basso2021optimized} &Ultrascale+  &22118 &4920 &0 &0 &250  \\  \hline 
\textbf{FIR.PolyMult} &Ultrascale+  &16971 &8755 &0    &0    &250  \\ \hline
\textbf{Fast.2.PolyMult } &Ultrascale+  &25831 &12850 &0 &0 &250 \\ \hline
\textbf{Fast.4.PolyMult} &Ultrascale+  &35306 &19143 &64 &0 &250  \\  \hline
Mera~\cite{mera2020compact} &Artix-7  &7400 &7331 &38 &2 &125  \\ \hline
\textbf{FIR.PolyMult} &Artix-7  &16902 &8755 &0    &0    &133   \\ \hline
\textbf{Fast.2.PolyMult} &Artix-7  &25854 &12850 &0 &0 &133 \\ \hline
\textbf{Fast.4.PolyMult} &Artix-7  &35396 &19143 &64 &0 & 133 \\ \hline       
\end{tabular}
}
\end{table*}

\begin{table*}[htbp]
  \centering
\caption{Timing performance of modular polynomial multiplier when $n=256$ in medium security level of Saber }
  \label{tb_mul_time}
   \begin{threeparttable}[htbp]{
\begin{tabular}{|c||c|c|c|c|c|c|}
\hline
Design  &Device  &1 PolyMult.$^a$  &KeyGen$^a$ &Encaps$^a$ &Decaps$^a$  &ATP-LUT$^b$\\ \hline 
Roy (1 Mult.)~\cite{roy2020high}$^c$   &Ultrascale+  &256 (1.02) &2685 (10.74) &3592 (14.37) &4484 (17.94) &7.49 $\times 10^5$   \\  \hline 
Roy (2 Mult.)~\cite{roy2020high}$^c$   &Ultrascale+ &128 (0.51) &1552 (6.21) &2205 (8.82) &2911 (11.64)  &8.50 $\times 10^5$ \\  \hline 
Zhu~\cite{zhu2021lwrpro} &Ultrascale+  &81 (0.81) &(Not Reported) &978 (9.78) &1227 (12.27)   &-\\ \hline
Basso (HS-I 256)~\cite{basso2021optimized}$^c$  &Ultrascale+  &256(1.02) &2685 (10.74) &3592 (14.37) &4484 (17.94) &4.67 $\times 10^5$  \\  \hline 
Basso (HS-I 512)~\cite{basso2021optimized}$^c$   &Ultrascale+ &128 (0.51) &1552 (6.21) &2205 (8.82) &2911 (11.64)   &5.89 $\times 10^5$ \\  \hline 
\textbf{FIR.PolyMult} &Ultrascale+  &511 (2.04) &2560 (10.24) &3328 (13.31)   &4096 (16.38)     &6.77 $\times 10^5$  \\ \hline
\textbf{Fast.2.PolyMult} &Ultrascale+   &255 (1.02) &1281 (5.12) &1665 (6.66) &2049 (8.20) &5.16 $\times 10^5$  \\ \hline
\textbf{Fast.4.PolyMult} &Ultrascale+   &127 (0.51) &642 (2.57) &834 (3.34) &1026 (4.10)  &3.50 $\times 10^5$ \\ \hline
Mera~\cite{mera2020compact} &Artix-7   &1299 (10.30) &11592 (92.74) &15456 (123.65) &19320 (154.56) &27.45 $\times 10^5$ \\  \hline 
\textbf{FIR.PolyMult} &Artix-7  &511 (3.83) &2560 (19.25) &3328 (25.02)    &4096 (30.80)    &12.66 $\times 10^5$  \\ \hline
\textbf{Fast.2.PolyMult} &Artix-7  &255 (1.92) &1281 (9.63) &1665 (12.52) &2049 (15.41) &9.71 $\times 10^5$ \\ \hline
\textbf{Fast.4.PolyMult} &Artix-7  &127 (0.95) &642 (4.83) &834 (6.27) &1026 (7.71)  &6.65 $\times 10^5$ \\ \hline
\end{tabular}
}
\begin{tablenotes}
     \item  {$^a$: Total latency in the unit of clock cycle (actual latency in the unit of $\mu$s) of one modular polynomial multiplication, or all the modular polynomial multiplications in Saber's specific step}
     \item  {$^b$: ATP-LUT (area-time product of LUTs) is calculated from the number of LUTs times the sum of actual latency ($\mu$s) of the total number of modular polynomial multiplications in KeyGen, Encaps, and Decaps steps}
     \item  {$^c$: Clock cycles for reading and writing operations are not counted }
   \end{tablenotes}
 \end{threeparttable}
 \vspace{-1em}
\end{table*}

The performance of the proposed modular polynomial multiplication is demonstrated for the Saber scheme using Verilog HDL implementation. Several changes have been adopted specifically for the Saber scheme. Due to the Saber scheme's advantages, the basic components do not consume a large amount of hardware resources. In particular, the modular multiplier discussed in \secref{sec_fir} can be replaced by a few adders since the coefficients of polynomial $B(x)$ are in the range of $[-4,4]$~\cite{d2018saber}. As the moduli are power-of-two integers, the modular reduction can again be performed by simply keeping the 
lower bits. Note that, the coefficients in both polynomials $A(x)$ and $B(x)$ are represented in the sign-magnitude form, and the word-lengths of the magnitudes of these two polynomials are 13-bit and 3-bit, respectively. 
The modular adder calculates the 13-bit sum ($sum$) by adding the product ($prod$) of the corresponding $a[i]$ and $b[j]$, and the output from the register $acc$ as shown in \figref{sequence_archi}(b), which can also be mathematically expressed as: 
\begin{equation}
       sum = 
       \begin{cases}
       acc-prod, & \text{if } a_{sign} \bigoplus b_{sign} = 1, \\
           acc+prod, & \text {otherwise, }
       \end{cases}
    \eqnlbl{modadd}
\end{equation}
where $a_{sign}$ and $b_{sign}$ are the sign bits of the two operands $a[i]$ and $b[j]$, respectively. Note that all the modular polynomial multiplications correspond to degree $n=256$.

The experiment is performed on the Xilinx Artix-7 AC701 FPGA since Artix-7 family FPGAs are recommended by NIST for PQC hardware implementation.
In addition, since several prior works also used the high-performance Xilinx UltraScale+ FPGA, we also demonstrate the performance of our architecture on this FPGA for more comparisons. 
The communication and data transmission between FPGA and PC use the universal asynchronous receiver transmitter (UART) module provided by the AC701 device for functionality verification.

\subsection{Evaluation of modular polynomial multiplier}\label{sec_mul_per}
We first examine the performance of our proposed modular polynomial multipliers, including the FIR filter-based (\figref{sequence_archi}), fast $2$-parallel architecture, and fast $4$-parallel architecture.

The experimental results and comparison with prior works~\cite{mera2020compact,roy2020high,zhu2021lwrpro, basso2021optimized} are summarized in Table~\ref{tb_mul}. A further comparison of the timing performance is presented in Table~\ref{tb_mul_time}. 
The clock frequencies are set as 250MHz and 133MHz for UltraScale+ and Artix-7, respectively. 

Table~\ref{tb_mul_time} summarizes the number of clock cycles and actual latency for one modular polynomial multiplication (PolyMult.), and all the modular polynomial multiplications in KeyGen, Encaps, and Decaps steps of Saber scheme with the medium security level. 
Note that the number of modular polynomial multiplications in encryption (decryption) is the same as in Encaps (Decaps). 
It can be seen from Table~\ref{tb_mul} that our design has a shorter critical path than those of the designs in~\cite{zhu2021lwrpro,mera2020compact} and the same as the work in~\cite{roy2020high,basso2021optimized}.

For a fair comparison, we focus on the evaluation against the architecture~\cite{roy2020high}, and its extended work~\cite{basso2021optimized} since both works use the same clock frequency and target high-speed design. 

Note that the high-speed designs in~\cite{basso2021optimized} do not provide the timing performance in the KeyGen, Encaps, and Decaps steps for Saber scheme but only the result for one modular polynomial multiplication; so we adopt the number of clock cycles from their previous work~\cite{roy2020high} and present in Table~\ref{tb_mul_time}. 
In particular, the framework and the timing performance (including the number of clock cycles and frequency) for one modular polynomial multiplication of~\cite{basso2021optimized} are maintained to be the same as their previous work~\cite{roy2020high}, while their optimized centralized multiplier design for the MAC unit in~\cite{basso2021optimized} significantly reduces LUTs. Besides, these two works present the general design and parallel design. In Tables~\ref{tb_mul} and~\ref{tb_mul_time}, the general designs correspond to Roy (1 Mult.) as well as Basso (HS-I 256), and parallel designs correspond to Roy (2 Mult.) as well as Basso (HS-I 512).

Compared to the general design in~\cite{roy2020high}, our proposed FIR filter-based modular polynomial multiplier (i.e., FIR.PolyMult) has slightly fewer LUTs and a smaller total latency in the modular polynomial multiplications used in three steps of the Saber scheme; however, a higher number of flip-flops (FFs) is needed due to the additional shift registers. 
Note that the clock cycles used in one modular polynomial multiplication in Roy (1 Mult.) are only 256 clock cycles due to the fact that their result does not count the number of clock cycles used for reading and writing operations~\cite{basso2021optimized}, which is same as the response time in our proposed FIR.PolyMult design. When compared to Roy (2 Mult.), our fast $2$-parallel architecture achieves $18.91\%$, $25.08\%$, and $39.88\%$ reduction on the number of LUTs, latency, and area-time product of LUTs (ATP-LUT), respectively. 

\begin{table*}[htbp]
  \centering
  \caption{Performance of modular polynomial multiplier using fast $M$-parallel architecture when $n=180$ based on Artix-7 FPGA family}\label{tb_para}
 \begin{threeparttable}

  {
\begin{tabular}{|c||c|c|c|c|c|c|c|c|}
\hline
Design   &LUTs &FFs &DSPs  &Freq. [MHz] &1 PolyMult.$^a$ & 9 PolyMult.$^a$ &ATP-LUT$^b$ &Throughput$^c$\\  \hline 
\textbf{Fast.2.PolyMult }   &17902 &9096 &0  &133 &181 (1.36) &901 (6.77) &1.21 $\times 10^5$ &2\\ \hline
\textbf{Fast.3.PolyMult }  &21729 &11996 &60  &133 &122 (0.91) &602 (4.53)&0.98 $\times 10^5$ &3\\ \hline
\textbf{Fast.4.PolyMult }   &25110 &13633 &45  &133 &92 (0.69)&452 (3.40) &0.85 $\times 10^5$ &4\\ \hline
\end{tabular}
}   
\begin{tablenotes}
     \item  {$^a$: Total latency in the unit of clock cycle (actual latency in the unit of $\mu$s)}
     \item  {$^b$: ATP-LUT (area-time product of LUTs) is calculated from the number of LUTs times the sum of actual latency ($\mu$s) of nine modular polynomial multiplications}
     \item  {$^c$: Throughput in the unit of samples per clock cycle}
   \end{tablenotes}
   \vspace{-1em}
 \end{threeparttable}
\end{table*}

Besides, their extended work~\cite{basso2021optimized} is also taken into comparison. Our FIR.PolyMult and fast $2$-parallel architectures require a larger number of LUTs compared to their general design (Basso (HS-I 256)) and parallel design (Basso (HS-I 512)), but the latency of our two proposed designs is smaller than theirs. In this case, the performance of ATP-LUT is utilized for a fair comparison. The Basso (HS-I 256) design has a $31.02\%$ reduction compared to our FIR.PolyMult architecture. However, the ATP-LUT product of our fast $2$-parallel architecture is reduced by $13.24\%$ compared to the Basso (HS-I 512). This result implies that our fast $M$-parallel design has a superior performance compared to conventional parallel processing techniques.

Even though our design requires more FFs in the data-path and shift registers, we argue that it has a small influence on the overall performance of UltraScale+ and Artix-7 FPGAs since both devices have a much higher resource budget for FFs than LUTs. For example, our proposed FIR.PolyMult design consumes $12\%$ of LUTs (16971/134600), but only $3\%$ of FFs (8755/269200) in AC701 FPGA. Even for the fast 4-parallel architecture, only $7\%$ FFs are utilized. 

Furthermore, both modular polynomial multiplier in LWRpro~\cite{zhu2021lwrpro} and the compact modular polynomial multiplier in~\cite{zhu2021lwrpro,mera2020compact} use the Toom-Cook/Karatsuba algorithm with 8-level and 4-level, respectively. The compact polynomial multiplier in~\cite{mera2020compact} has a long critical path of five adders/subtractors and two multipliers in the interpolation part, which requires two pipelining stages to reduce the critical path for maintaining a high frequency. This design targets the low-area performance, which only requires limited numbers of LUTs, FFs, and only 38 DSP units, as shown in Table \ref{tb_mul}. 
While this design has lower LUT usage than our architecture, it suffers from a low speed since their length-$64$ polynomial multipliers require 1168 clock cycles for each computation, which causes the actual latency in such a compact design to be around 19 times of the latency in our fast $4$-parallel architecture as presented in Table \ref{tb_mul_time}. 
If we consider the ATP-LUT as the performance metric to compare our proposed fast $4$-parallel architecture and this prior low-area design, it shows that our design achieves a $75.77\%$ reduction. 

Besides, the modular polynomial multiplier in~\cite{zhu2021lwrpro} requires the lowest number of clock cycles among all the prior works, while having the lower clock frequency as illustrated in Table~\ref{tb_mul_time}. 
In comparison, our fast $4$-parallel architecture requires $15.65\%$ fewer clock cycles and achieves a $66.26\%$ reduction in the actual latency for all the modular polynomial multiplications in the Encaps and Decaps steps. Considering the area performance of this work, their modular polynomial multiplier uses $60.48\%$ fewer LUTs but $24.71\%$ more DSPs compared to the proposed fast $4$-parallel architecture. Thus, the ATP products of DSP and LUT need to be considered separately. Since the clock cycles used for KeyGen are not reported in~\cite{zhu2021lwrpro}, the ATP-LUT (ATP-DSP) for the comparison with this work is defined as the number of LUTs (DSPs) times the sum of actual latency ($\mu$s) of the total number of modular polynomial multiplications in two steps only, Encaps and Decaps. Specifically, the ATP-LUT and ATP-DSP in their modular polynomial multiplier are $3.08 \times 10^5$ and $1874.20$, respectively. The ATP-LUT and ATP-DSP in the fast $4$-parallel architecture are $2.63 \times 10^5$ and $476.16$, respectively. Under this comparison, ATP-LUT and ATP-DSP are reduced by $14.61\%$ and $75.07\%$, respectively, in our design. 

Thus, we can conclude that our design achieves a significant reduction in the latency or the delay (critical path), which leads to reductions in ATP compared to the two prior works that employ the Karatsuba/Toom-Cook algorithm-based modular polynomial multiplication.

Our proposed modular polynomial multipliers can be sufficient to support different security levels of Saber without any change of the area consumption and the frequency presented in Table~\ref{tb_mul}, but only requires a different number of clock cycles.

\subsection{Parallel architectures}
The works in~\cite{roy2020high} and~\cite{basso2021optimized} also present a parallel architecture, which is a scaled version. The parallel design in~\cite{roy2020high} (Roy (2 Mult.)) uses two multipliers in one MAC unit, and the parallel design in~\cite{basso2021optimized} (Basso (HS-I 512)) doubles their MAC units (each MAC unit has one MUX and one adder). When compared to the Roy (2 Mult.) design, our fast $2$-parallel architecture achieves a significant reduction in the area overhead and latency. In particular, our fast $2$-parallel architecture consumes only about $34\%$ higher area consumption than the FIR filter polynomial multiplier while reducing latency by $50\%$, while their scaled version of the parallel modular polynomial multiplier has $45\%$ overhead compared with the general design architecture (Roy (1 Mult.)). Along with parallelization, the delay also increases, as the critical path will change from one multiplication and one addition to one multiplication and two additions. In this case, an additional pipeline is added in the design of Roy (2 Mult.)~\cite{roy2020high} to maintain the same high frequency. Under the same number of pipelining stages, our fast $2$-parallel architecture achieves a lower critical path and hence can be driven by a clock with a higher frequency. 

\begin{table*}[htbp]
\setlength{\tabcolsep}{4pt}
    \centering
        \caption{Comparison with recent Saber scheme implementation in medium security level }\label{tb_overall}
        \begin{threeparttable}
  \setlength{\tabcolsep}{1pt}
\begin{tabular}{|c||c|c|c|c|c|c|c}
\hline
    & Platform&\shortstack{Time in 
    ($\mu s$): KeyGen/Encaps/Decaps}&\shortstack{Freq. [MHz]} & \shortstack{Area: LUTs/FFs/DSPs/BRAM} &ATP-LUT$^a$ \\
     \hline 
    
    Roy~\cite{roy2020high}&UltraScale+ &21.8/26.5/32.1&250&23.6k/9.8k/0/2&1.9 $\times 10^6$ \\\hline
    \textbf{Ours}    &UltraScale+&10.2/12.6/15.6&250&41.5k/22.3k/64/2 &1.6 $\times 10^6$ \\ \hline
    \textbf{Ours}    &Artix-7&19.2/23.6/29.2&133&41.5k/22.3k/64/2 &3.0 $\times 10^6$ \\ \hline
    
    \end{tabular}
    
    \begin{tablenotes}
     \item  {$^a$: \small{ATP-LUT (area-time product of LUTs) is calculated from the number of LUTs times the sum of actual latency ($\mu$s) of KeyGen, Encaps, and Decaps steps}}
     \vspace{-1em}
   \end{tablenotes}
 \end{threeparttable}
\end{table*}

We also compare different fast $M$-parallel architectures for $n=180$ in Table \ref{tb_para}. It can be noticed that when the level of parallelism increases ($M$ becomes larger), the actual latency is reduced at the expense of higher area consumption. Besides, the throughput of the designs is increased when $M$ becomes larger. The ATP-LUT product for the fast $2$-parallel, $3$-parallel, and $4$-parallel architectures are listed for computing nine modular polynomial multiplications, which indicates that a higher level of parallelism can provide a more efficient design if sufficient resource budget is available.

\subsection{Comparison with Saber PQC scheme implementations}\seclbl{resu_saber}
For the implementation of the entire Saber scheme, the modular polynomial multiplication is implemented by the proposed fast $4$-parallel architecture, while other simple functional blocks are modified from the open-source codes provided in~\cite{d2018saber} and~\cite{roy2020high}.

Table~\ref{tb_overall} presents the comparison of the FPGA performance with recent hardware implementation~\cite{roy2020high} for the Saber PQC scheme at a medium security level. The latency in our design is $52\%$ less than the latency in~\cite{roy2020high} with the cost of more LUTs and DSPs consumed. In fact, the reduction is mainly from our optimized low-latency modular polynomial multiplier. Besides, instead of directly adopting the open-source SHA3 hash function block as in~\cite{roy2020high}, we also implement the hash function block when implementing the entire scheme. For example, the total latency of SHA3-256 (needs to process 32-byte, 64-byte, 992-byte, and 1088-byte seeds) operating in the hash function block is reduced from 585 clock cycles to 526 clock cycles in the Saber Encaps. The rationale behind this latency reduction is as follows. Most open-source packages add stages of pipelining to achieve a high frequency (low critical path) design in order to adapt to general applications~\cite{sundal2017efficient}. However, the critical path of the modular polynomial multiplier that requires addition or multiplication from prior work is much higher than the Keccak core provided in the open-source packages, thus implying that some pipelines are redundant. Different from the prior work, we implement our own hash function block as we aim to reduce the total latency for computing the hash functions by eliminating unnecessary pipelining stages. 

For the area performance, although we have increased hardware costs, both Artix-7 and UltraScale+ FPGAs still have sufficient resources to accommodate our fast $4$-parallel design. In other words, our proposed fast $4$-parallel architecture is under the constraint of hardware complexity specified by NIST (Artix-7).

\section{Conclusion}\seclbl{con}
This paper has presented a novel modular polynomial multiplier and demonstrated its applications for lattice-based cryptography. The proposed hardware design exploits the fast filtering technique to achieve low latency, high scalability, and full hardware utilization. We proposed efficient parallel architectures with much lower hardware overhead and latency than prior works. Our design can be easily generalized across different levels of parallelism. Comprehensive experimental results are presented. We show that our design achieves superior performance than the state-of-the-art modular polynomial multipliers based on schoolbook polynomial multiplication or the Karatsuba algorithm. A case study of the implementation of the Saber scheme shows that our proposed design can accelerate the computation and reduce the actual latency of the cryptosystem compared with the prior work.

\section{Disclosure}
Part of this paper is covered in the patent application~\cite{parhi2022low}.

\bibliographystyle{IEEEtran}
\bibliography{main}

\begin{IEEEbiography}[{\includegraphics[width=1in,height=1.25in,clip]{TAN.pdf}}]{Weihang Tan} (Graduate Student Member, IEEE) is a postdoctoral research associate in the Department of Electrical and Computer Engineering at University of Minnesota. He received his B.S., M.S., and Ph.D. degrees in Electrical Engineering from Clemson University, Clemson, SC, USA, in 2018, 2020, and 2022, respectively. His research interests include hardware security and VLSI architecture design for fully homomorphic encryption, post-quantum cryptography, and digital signal processing systems. He is the recipient of the best PhD forum presentation award at the Asian Hardware Oriented Security and Trust Symposium (AsianHOST).
\end{IEEEbiography} 

\begin{IEEEbiography}[{\includegraphics[width=1in,height=1.25in,clip,keepaspectratio]{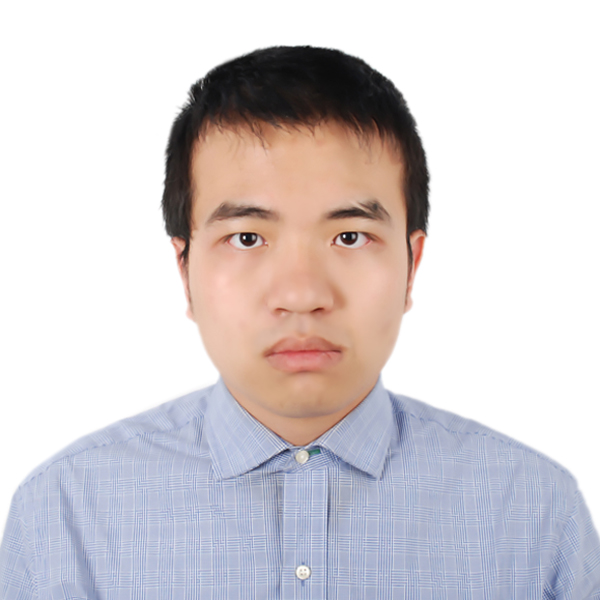}}]{Antian Wang} (Graduate Student Member, IEEE) 
received his B.E. (2017) in Communication Engineering from Shanghai Maritime University. He is currently pursuing a Ph.D. degree in Clemson University. His research interests include hardware security and VLSI architecture design, and design automation.
\end{IEEEbiography} 

\begin{IEEEbiography}[{\includegraphics[width=1in,height=1.33in,clip]{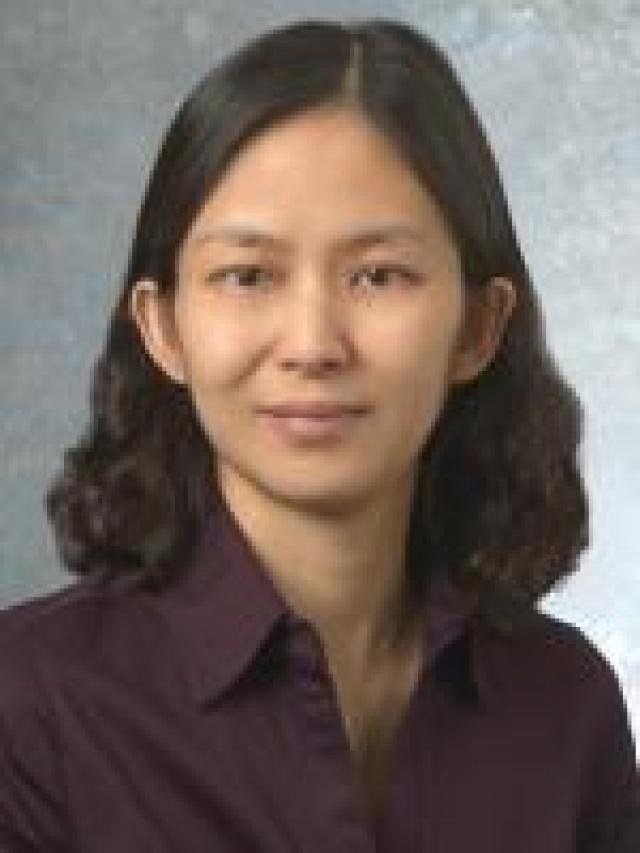}}]{Xinmiao Zhang} (Senior Member, IEEE) Received her Ph.D. degree from the University of Minnesota. She joined The Ohio State University as an Associate Professor in 2017. Prior to that, she was a Senior Technologist at Western Digital, an Associate Professor at Case Western Reserve University. Her research interests include VLSI architecture design, digital storage and communications, security, and signal processing. She published more than 100 papers and authored the book "VLSI Architectures for Modern Error-Correcting Codes" (CRC Press, 2015). Dr. Zhang is a recipient of the NSF CAREER Award in 2009 and the Best Paper Award at 2004 ACM Great Lakes Symposium on VLSI. She was elected to serve on the BoG (2019-2021) of the IEEE CASS. She is also the Chair (2021-2022) of the Data Storage Technical Committee. She served on the committees of many conferences, including ISCAS, SiPS, ICC, and GLOBECOM. 
\end{IEEEbiography}

\begin{IEEEbiography}[{\includegraphics[width=1in,height=1.25in,clip]{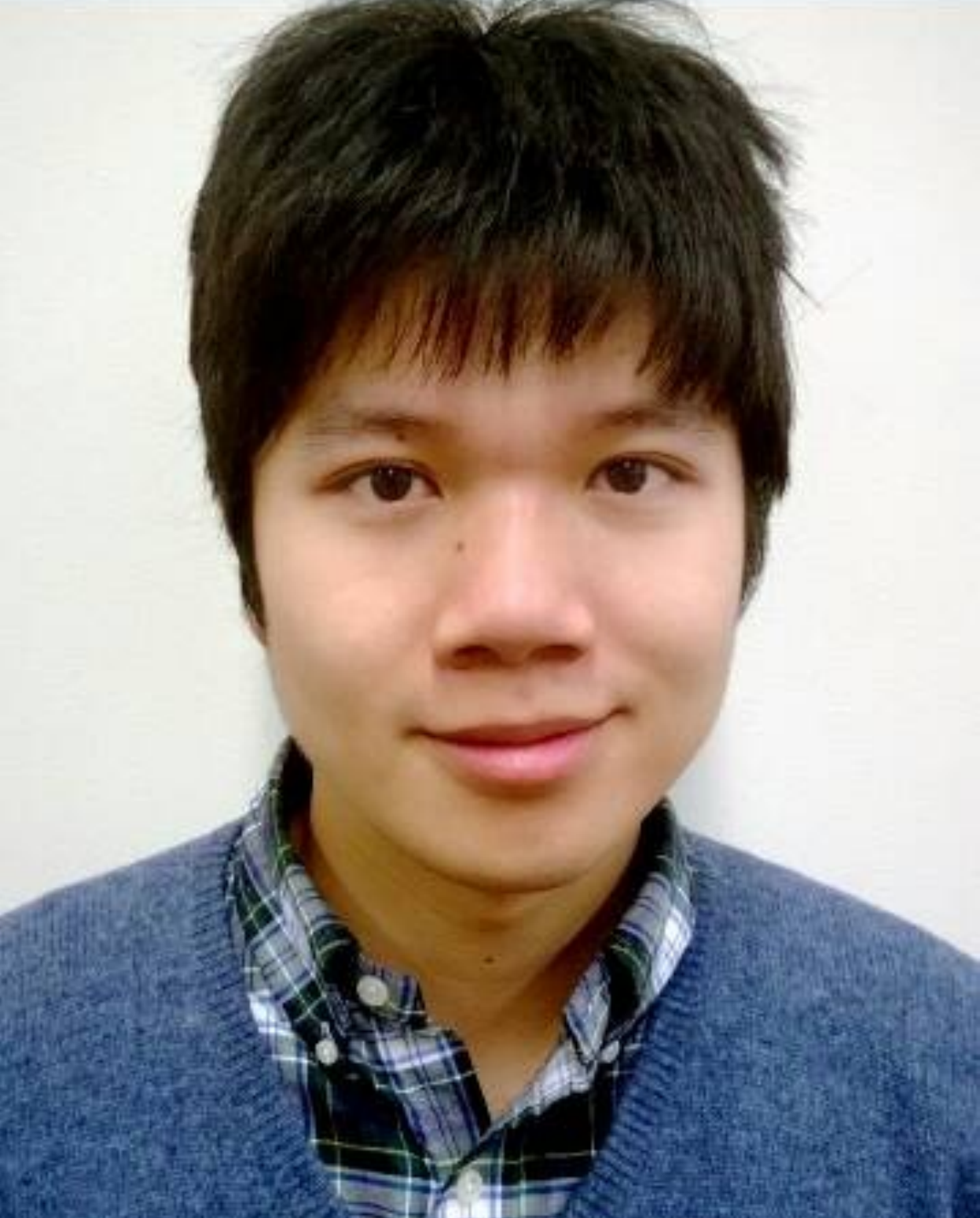}}]{Yingjie Lao} (Senior Member, IEEE) is currently an assistant professor in the Department of Electrical and Computer Engineering at Clemson University. He received the B.S. degree from Zhejiang University, China, in 2009, and the Ph.D. degree from the Department of Electrical and Computer Engineering at University of Minnesota, Twin Cities in 2015. He is the recipient of an NSF CAREER Award, a Best Paper Award at the International Symposium on Low Power Electronics and Design (ISLPED), and an IEEE Circuits and Systems Society Very Large Scale Integration Systems Best Paper Award. 
\end{IEEEbiography}

\begin{IEEEbiography}[{\includegraphics[width=1in,height=1.16in,clip]{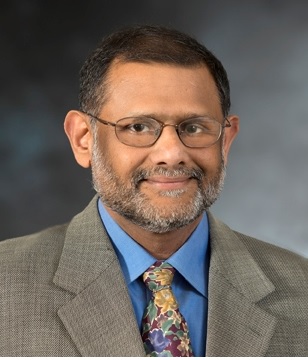}}]{Keshab K. Parhi} (Fellow, IEEE) is Distinguished McKnight University Professor and Erwin A. Kelen Chair Professor in the Department of Electrical and Computer Engineering. He completed his Ph.D. in EECS at the University of California, Berkeley in 1988. He has published over 700 papers, is the inventor of 34 patents, and has authored the textbook VLSI Digital Signal Processing Systems (Wiley, 1999). His current research addresses VLSI architectures for machine learning, hardware security, and data-driven neuroscience. Dr. Parhi is the recipient of numerous awards including the 2003 IEEE Kiyo Tomiyasu Technical Field Award, and the 2017 Mac Van Valkenburg award and the 2012 Charles A. Desoer Technical Achievement award from the IEEE Circuits and Systems Society. He served as the Editor-in-Chief of the {\em IEEE Trans. Circuits and Systems, Part-I: Regular Papers} during 2004 and 2005. He is a Fellow of the ACM, AIMBE, AAAS, and NAI.
\end{IEEEbiography}

\end{document}